\documentclass[useAMS,usenatbib]{mn2e}
\usepackage{float,graphics,epsfig,array,amsmath,amssymb,subfigure}

\title[Bayesian Photometric Redshifts]
{Bayesian Photometric Redshifts for Weak Lensing Applications}
\author[E.M. Edmondson et al. ]{E.M. Edmondson$^{1,2}$, L. Miller$^1$,
  C. Wolf$^1$
  \\  $^1$ Department of Physics, Oxford University, Keble Road, Oxford,
  OX1 3RH, U.K.
  \\ $^2$ Institute of Cosmology and Gravitation, University of
  Portsmouth, Portsmouth, PO1 2EG, U.K.
}

\begin{document}
\topmargin -0.5in
\pagerange{\pageref{firstpage}--\pageref{lastpage}} \pubyear{2005}

\maketitle

\label{firstpage}

\begin{abstract}
The next generation of weak gravitational lensing surveys is capable
of generating good measurements of cosmological parameters, provided
that, amongst other requirements, adequate redshift information is
available for the background galaxies that are measured.  It is
frequently assumed that photometric redshift techniques provide the
means to achieve this.  Here we compare Bayesian and frequentist
approaches to photometric redshift estimation, particularly at faint
magnitudes.  We identify and discuss the biases that are inherent in
the various methods, and describe an optimum Bayesian method for
extracting redshift distributions from photometric data.
\end{abstract}


\section{Introduction}
Photometric redshifts have been used since the 1960s~\citep{Baum} as a
means of estimating redshifts in surveys where it would be impractical
to obtain spectroscopic redshifts for all the objects observed, or
where objects are too faint for spectroscopic techniques to be
applied. Since then multicolour surveys have commonly used some form
of redshift estimation with varying degrees of success depending on
the selection of filters used and the method of relating particular
sets of colours to a redshift estimate.

The importance of the technique is growing not only with the desire to
gain a greater understanding of galaxy evolution (through, for
example, the determination of luminosity functions) but also in weak
gravitational lensing, where redshift estimates can reduce
contamination from intrinsic
alignments~\citep{HeavensHeymans,KingSchneider}, and allow the
possibility of 3D lensing studies.  \citet{Hu} and \citet{Heavens}
proposed 3D lensing analyses that provide better constraints on the
mass power spectrum and vacuum energy equation of state than
traditional `2D' analyses. \citet{Taylor} discussed a method of using
redshift information to reconstruct the 3D lensing potential, and this
has been applied to simulations~\citep{BaconTaylor} and the COMBO-17
multi-waveband survey~\citep{TaylorBacon}.

In all these applications the aim is to estimate redshifts photometrically
at magnitudes fainter than may readily be achieved by multi-object
spectroscopy.  In the case of weak lensing studies in particular, there
is a strong motivation to measure the lensing signal at the faintest
possible magnitudes.  But of course at faint magnitudes the photometric
measurement errors become significant and cause increased 
redshift errors.  We shall show below that this leads to a bias
in the estimated redshift distribution obtained which for 3D weak lensing
and galaxy evolution studies could seriously affect the results.
In recognition of the difficulty of obtaining reliable photometric redshifts
at faint magnitudes, it is common instead to assume a statistical distribution
for redshifts which may be calculated given knowledge of the evolving
galaxy luminosity function.

In this paper we discuss a Bayesian approach to redshift estimation,
examining specifically the case of a survey utilising $UBVRI$ imaging
since large area weak lensing surveys will be expected to use such
broad-band datasets.  We show that, by adopting a prior calculated
from galaxy luminosity functions, it is possible both to correct for
the bias in the sample distribution and to obtain redshift
distributions that smoothly converge to the prior distribution at the
limit of faint magnitudes.  We show however that there is a price that
must be paid: for each galaxy it becomes necessary not to assign a
single definite redshift, but rather to consider its entire posterior
probability distribution in redshift, in order to avoid bias.

The idea of applying a prior to improve photometric redshift estimates
is not new. \citet{Benitez} has discussed adopting a Bayesian analysis
in photometric redshift analysis and tested this using a parameterised
form for the prior on a spectroscopic sample in the Hubble Deep
Field North.  Some improvement in redshift error
was obtained, and by inspecting the values of the posterior probability
distribution, outliers in the distribution of sample redshifts 
could be removed.  

\citet{CFDF-PRS} used an iterative method to estimate a prior that
improved the estimate of the overall redshift distribution, but this
did not follow a strictly Bayesian approach and the method was shown
to break down at very low signal-to-noise.

In fact, consideration of Bayes' theorem
leads us to expect that it should be possible to calculate a prior which
is indeed based on our prior knowledge, and that the application of this
to the redshift estimation problem should not require iteration or training.
Furthermore, a full Bayesian approach requires us to consider the
complete posterior probability distribution, and we discuss in the next
section why this is necessary to avoid biased distributions at faint
magnitudes.

\section{The Bayesian method}

\subsection{Introduction to the Bayesian method}

There are two classes of approach to the problem of photometric
redshift estimation. One is to fit colours by a range of templates (as
employed by Hyperz and CFDF-PRS - see \citealt{hyperz} and
\citealt{CFDF-PRS} respectively) and the second is to utilise some
method of machine learning such as the artificial neural networks of
ANNz~\citep{collister04}, support vector machines (a type of learning
algorithm for general classification problems) as used
by~\citet{wadadekar}, or polynomial fits to a spectroscopic sample
(\citealt{connolly}, and later ~\citealt{hsieh}). Whilst the
second approach as demonstrated by ANNz is highly effective for
reasonably bright objects where it is practical to obtain a large
training set of spectroscopically identified objects, we use methods
based on the former approach as they are more readily used when such
training sets are not available, as is the case for samples of faint
objects.

Suppose we have some measured colour information, denoted by $C$, for
a galaxy.  Traditional template-based photo-z methods work with the
likelihood $\mathcal{L}(C|z)$.  A Bayesian approach is to consider
instead the posterior probability $p(z|C,\mathcal{P}) \propto
\mathcal{L}(C|z) p(z|\mathcal{P})$ where $\mathcal{P}$ denotes our
prior knowledge and $p(z|\mathcal{P})$ is the corresponding prior
probability distribution for $z$.  The source of prior information we
wish to use here is the galaxy luminosity function, $\phi$, determined
from spectroscopic redshift surveys.  In order to use this information
we need to determine the redshift distribution that results from a
known luminosity function as a function of galaxy magnitude $m$ and
spectral type $S$: $p(z|\mathcal{P}) = p(z| m, S, \phi(m,S,z) )$.
This prior distribution is a more general form of the redshift
distribution that has previously been used in the limiting case of
assuming no colour information (e.g. \citealt{BrownTaylor} who assume a
distribution based on extrapolating the median redshift from magnitude
ranges where photometric redshifts are available out to fainter
magnitudes where they are not), and hence the Bayesian approach can be
seen to fully encapsulate the transition between the high
signal-to-noise regime, where $\mathcal{L}(C|z)$ is sharply defined
and the prior has no effect, and the low signal-to-noise regime where
the prior information dominates.

\subsection{Noise bias and choice of prior}
\label{choiceofprior}
Before discussing the application of such a prior, we first need to highlight
the bias that is introduced by using only a likelihood-based estimator in
the presence of noise, and also to discuss the extent to which applying 
a Bayesian prior may alleviate that bias.  

Consider a sample of galaxies with a true redshift distribution $n(z)$
and suppose those redshifts are measured by any technique which has
some error distribution $\varepsilon(z_o | z)$ where $z_o$ and $z$ are
the measured and true redshifts respectively.  Then the observed
sample distribution of redshifts is the convolution $n(z) \ast
\varepsilon(z_o | z)$ which is different from, and typically broader
than, the true redshift distribution.  The difference between true and
measured distributions is small at high signal-to-noise where
$\varepsilon(z_o | z)$ tends to a delta function, but becomes
significant when the width of the $\varepsilon(z_o | z)$ distribution
is comparable to the width of the true distribution.

A Bayesian approach circumvents this problem.  In practice we measure a set
of photometric data $C$, with probability of obtaining that
data $\varepsilon(C | z)$.  We assume a likelihood function
$\mathcal{L}(C | z)$, where $\mathcal{L}(C | z) \equiv \varepsilon(C | z)$
if the likelihood function is a good representation of the true
process by which the observed galaxy data is created.
We can calculate a normalised Bayesian
posterior probability distribution 
\begin{equation}
p(z | C, \mathcal{P}) = \frac{ \mathcal{P}(z) \mathcal{L}(C | z)}
                               { \int dz \mathcal{P}(z) \mathcal{L}(C | z)}.
\end{equation}
But the photometric distribution of galaxies at this measured redshift
is the convolution of the true distribution with the error
distribution as discussed above, and hence the sum of the normalised
posterior probability distribution for the sample is
\begin{eqnarray}
\nonumber
\lefteqn{\int dC n(C) p(z | C, \mathcal{P}) = } \\
\nonumber
& & \int dC \frac{ \mathcal{P}(z) \mathcal{L}(C | z)}{ \int dz \mathcal{P}(z) \mathcal{L}(C | z)}
\int dz' n(z') \varepsilon(C | z')\\
& = & n(z)
\end{eqnarray}
if $\mathcal{L}(C | z) = \varepsilon(C | z)$ and we choose
$\mathcal{P}(z) = n(z)$.  Hence the posterior probability distribution
may be used to create sample redshift distributions which are unbiased
provided
\begin{enumerate}
\item the likelihood function is an accurate reflection of the true statistical
process (this requirement also applies to likelihood-based methods of course)
\item the prior is sufficiently well-known. 
\end{enumerate}

\subsection{Choice of estimator}
\label{choiceofestimator}
From the above, we can see that a likelihood-based estimator (either
maximum likelihood or minimum variance) should give adequate results
in the limit of high signal-to-noise where the effect of ignoring the
prior is unimportant.  We shall argue below that at faint magnitudes,
however, the Bayesian posterior probability distribution should be
used.  This makes the process of utilising photometric redshifts at
faint magnitudes more problematic, as we now no longer have a single
estimated value, perhaps with some quoted uncertainty, but instead
have a continuous probability distribution.

One approach would be to replace, say, a maximum likelihood estimator
by a maximum posterior probability (MPP) estimator or similar, as
advocated by \citet{Benitez}.  Such an estimator, which would simply
return $\max p(z)$ rather than $\max \mathcal{L}(z)$ would still be
biased, however, albeit differently from the ML estimator.  In the
limit where the prior dominates over the likelihood function, the MPP
estimator would place all galaxies at the same redshift, at the peak
of the prior distribution.  So we can see that the ML estimator tends
to produce a sample distribution which is too broad, an MPP estimator
a sample distribution that is too narrow, with only the full posterior
probability distribution encapsulating the information in an unbiased
way.

These considerations lead to the supposition that there might exist an
estimator which would create an unbiased sample distribution, which
would be obtained by applying a prior somewhere between the uniform
case assumed in the ML estimator and the full Bayesian prior.  In the
case of a simple distribution, such as the normal distribution, it is
straightforward to calculate such a prior.  For more complex
distributions it may only be evaluated numerically.  We refer to this
as a ``sample prior'', but in light of the evalutation of the
alternative technique presented in this paper we do not believe that
this is a useful approach due to the difficulties involved in
correctly calculating such a prior from Monte Carlo simulations. This
technique is discussed in more detail in Section~\ref{sampleprior}.

\section{Application to a test dataset}

\subsection{The COMBO-17 survey}
To test whether the ideas discussed above are applicable to real
datasets, and to measure the difference in results that may be
obtained, we test the method on photometric data from COMBO-17
survey~\citep{COMBO-17} and spectroscopic redshifts obtained from
VVDS~\citep{VVDS}, with the two datasets overlapping on the Chandra
Deep Field South (CDFS). COMBO-17 has imaging in 17 optical filters
from the Wide Field Imager on the MPG/ESO 2.2m telescope sited at the
La Silla observatory in Chile, and reaches depths of $R<25.4$
($10\sigma$ limit for the CDFS, Vega magnitude).  The full 17-band
COMBO-17 photometric redshifts have accuracies of $\delta z/(1+z)
\approx 0.025$ for $R<23$ (approximately 2000 objects) rising to
$0.06$ for $R<24$ (10,000 objects).  VVDS is a spectroscopic survey of
11,564 objects over 0.6 square degrees using VIMOS on ESO-VLT UT3. The
sample is selected purely by $17.5\leq I_{AB} \leq 24$.

\begin{figure}
\resizebox{0.8\linewidth}{!}{\includegraphics{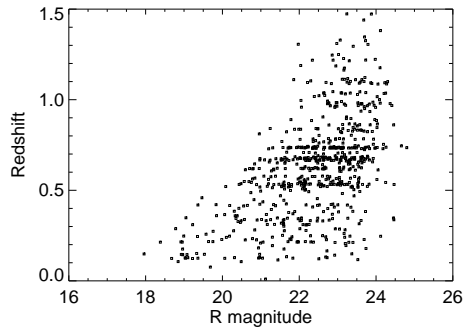}}
\caption{VVDS galaxy spectroscopic redshifts and magnitudes, limited
to $z<1.5$ (excluding 0.3 per cent of the sample)}
\label{vvds_distribution}
\end{figure}

\begin{figure}
\resizebox{0.8\linewidth}{!}{\includegraphics{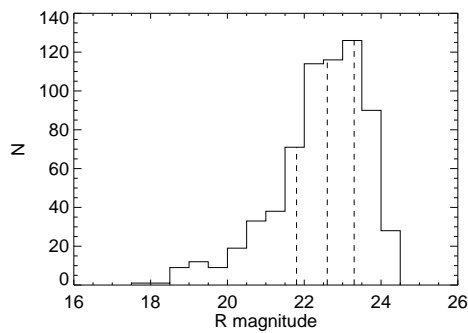}}
\caption{VVDS magnitude distribution, $R$-band aperture
magnitudes. 25th, 50th and 75th percentiles marked.}
\label{m_distribution}
\end{figure}

We use here a sample of 671 galaxies, selected to have photometry from
the CDFS field of COMBO-17 and spectroscopic redshifts from VVDS. We
select only VVDS galaxies with redshifts assigned such that more than
95 per cent are correct (corresponding to VVDS redshift flags of 3, 4, 23
or 24). Whilst lower-confidence VVDS redshifts would be acceptable in
many other applications where redshift errors might average out, we
choose to be highly conservative in order to ensure that our
assessment of photometric redshift errors are not distorted by errors
in our spectroscopic sample. The distribution of these galaxies in $R$
and $z$ is shown in Fig.~\ref{vvds_distribution}. The median $R$ of
the sample is 22.6, with 25th and 75th percentiles at $R=21.8$ and
$R=23.3$ (Fig.~\ref{m_distribution}).

Of the 671 VVDS galaxies 21 do not have colours close to those of a
template and produce problematic redshift estimates as a result, and
are therefore excluded. Two of the remaining objects have a true
redshift beyond $z=1.5$, and are also excluded from the set, bringing
the total removed to 3.4 per cent. Three further objects are at
$1.4<z<1.5$, and whilst strictly beyond the limit of the estimation
templates are quite close and are still included.

In addition, COMBO-17 provides a set of galaxy luminosity functions
for three broad galaxy types, derived from all three fields. These
types are separated by rest-frame $U-V$ colour-cuts evolving with
redshift, with a cut between blue and red populations and a further
cut between two subsets of the blue population (Table
~\ref{colour_cut_table}). An earlier variant of the calculation of
these luminosity functions is described in~\citet{COMBO-17GLF} and a
more recent calculation in~\citet{Faber}, which describes functions
much like those outlined here but where the two Blue types are
combined into a single set. The three types used here are separated by
rest-frame $U-V$ colour-cuts evolving with redshift, with a cut
between blue and red populations and a further cut between two subsets
of the blue population (Table ~\ref{colour_cut_table}). This data is
used in the construction of the prior itself (see
Section~\ref{constructing_prior}).

\begin{table}
\small{
\begin{tabular}{|c|c|c|c|}
\hline
Type & $z$ & $\phi^* / 10^4 \textrm{(Mpc h}^{-1})^3$ & $M^*$ \\
\hline
Blue 1 & 0.3 & $44.09 \pm 14.18$ & $-20.00 \pm 0.23$ \\
       & 0.5 & $36.79 \pm 6.97$ & $-20.45 \pm 0.18$ \\
       & 0.7 & $42.05 \pm 1.86$ & $-20.67 \pm 0.23$ \\
       & 0.9 & $58.81 \pm 20.74$ & $-20.48 \pm 0.17$ \\
       & 1.1 & $49.60 \pm 4.13$ & $-20.56 \pm 0.17$ \\
\hline
Blue 2 & 0.3 & $42.76 \pm 19.92$ & $-19.53 \pm 0.23$ \\
       & 0.5 & $37.30 \pm 9.98$ & $-20.00 \pm 0.18$ \\
       & 0.7 & $61.84 \pm 13.69$ & $-19.88 \pm 0.14$ \\
       & 0.9 & $74.51 \pm 25.50$ & $-19.87 \pm 0.14$ \\
       & 1.1 & $49.05 \pm 14.55$ & $-20.20 \pm 0.14$ \\
\hline
Red    & 0.3 & $63.87 \pm 24.71$ & $-19.86 \pm 0.16$ \\
       & 0.5 & $58.21 \pm 9.35$ & $-20.00 \pm 0.11$ \\
       & 0.7 & $51.76 \pm 2.05$ & $-20.33 \pm 0.12$ \\
       & 0.9 & $34.67 \pm 15.16$ & $-20.41 \pm 0.14$ \\
       & 1.1 & $15.52 \pm 3.48$ & $-20.81 \pm 0.16$ \\
\hline
\end{tabular}
\begin{tabular}{|c|c|}
\hline
Type & $\alpha$ \\
\hline
Blue 1 & $-1.3 \pm 0.1$ \\
Blue 2 & $-1.3 \pm 0.1$ \\
Red & $-0.5 \pm 0.1$ \\
\hline
\end{tabular}
}
\caption{Galaxy luminosity function parameters. $M^*$ is given in rest-frame Johnson B.}
\label{glf_table}
\end{table}

\begin{table}
\small{
\begin{tabular}{|c|c|c|}
\hline
$z$ & Blue 1 / Blue 2 cut & Blue 2 / Red cut \\
\hline
0 & 0.760 & 1.81 \\
0.25 & 0.592 & 1.66 \\
0.35 & 0.525 & 1.60 \\
0.45 & 0.416 & 1.55 \\
0.55 & 0.391 & 1.49 \\
0.65 & 0.239 & 1.41 \\
0.75 & 0.238 & 1.39 \\
0.85 & 0.149 & 1.40 \\
0.95 & 0.082 & 1.35 \\
1.05 & -0.045 & 1.34 \\ 
1.15 & 0.104 & 1.34 \\
1.25 & 0.045 & 1.30 \\
1.35 & -0.177 & 1.26 \\
\hline
\end{tabular}
}
\caption{Colour cuts between galaxy populations expressed in terms of
$(U-V)_\mathrm{rest}+0.08\times(M_V+20)$.}
\label{colour_cut_table}
\end{table}

\subsection{Estimation technique and construction of the Bayesian prior}
\subsubsection{Spectral models}
Galaxy spectral models are taken from PEGASE stellar population
synthesis models~\citep{PEGASE} as detailed in the COMBO-17 CDFS
reference paper~\citep{COMBO-17}.  By shifting the template
wavelengths and passing through the appropriate filter models colour
templates are obtained. In all 360 different spectral energy
distribution templates (SEDs) are shifted over 177 intervals
equidistant in $\log (1+z)$ running from $z=0$ to $z=1.4$. 

The redshift range is restricted to $z<1.4$ owing to a lack of high
contrast spectral features in the bands used (a problem that could be
alleviated by observing in the near infra-red), exacerbated
by noisy photometry in the typically faint $z>1.4$ objects, both of
which severely limit the ability to make accurate estimates.

Spectral models for stars and quasars (based on templates from
\citealt{Pickles} and \citealt{vandenberk} respectively) are also
used. A decision is made into which of these three classes an object
should fall by normalising the total likelihoods for each by the
number of templates available in each class, multiplying by an overall
class prior based on $I$-band magnitudes and choosing the class with
the highest normalised likelihood.

\subsubsection{Template fitting}
In order to best allow for the effect of errors in the measurement it
is preferable to perform a fit in terms of flux, rather than
magnitude, as the error is symmetric in the former but not in the
latter (for the case where background noise dominates - the low signal
regime at which the error becomes most significant for photometric
redshift estimation), and this asymmetry becomes important as the size
of the error term rises. A natural way to express the fit to a set of
observed fluxes $\mathbf{f}$ to a set of template fluxes $\mathbf{\hat{f}}$
would be
\begin{equation}
p(z,S|F,\mathcal{P}) = p(z,S|\mathcal{P}) \exp \left[ -\frac{1}{2}
  \left( \frac{\mathbf{f-\hat{f}}}{\boldsymbol{\sigma_f}}\right) ^2
  \right] .
\label{fluxlikelihood}
\end{equation}
However we do not have a set of template fluxes $\mathbf{\hat{f}}$ -
we only have a set of colours, which are equivalent to template flux
ratios. We can convert the above to an expression involving these flux
ratios if we arrange the ratios in terms of $\hat{f}_R$ as
$\hat{\lambda}_i = \hat{f}_i / \hat{f}_R$, and using $\mathbf{\hat{f}}
= \hat{f}_R \boldsymbol{\hat{\lambda}}$ and marginalising over all
$\hat{f}_R$ from $0$ to $\infty$ giving
\begin{equation}
p(z,S|f,\mathcal{P}) \propto \int^{\infty}_0 p(z,S|\mathcal{P}) \exp
\left[ -\frac{1}{2} \left( \frac{\mathbf{f}-\hat{f}_R
\boldsymbol{\hat{\lambda}}}{\boldsymbol{\sigma_f}} \right) ^2 \right]
d\hat{f}_R .
\label{likelihood}
\end{equation}
It is this equation that is evaluated to obtain our photometric
redshifts (although the integration can in practice be reduced to the
range where $\hat{f}_R$ is within a few $\sigma_f$ of $f_R$ since
$\hat{\lambda}_R = 1$ and the value of the integrand approaches zero
outside this region).

The resulting $p(z,S|f,\mathcal{P})$ can be marginalised over the SED
range $S$ to obtain a redshift estimate (e.g. by selecting the maximum
likelihood (ML) or minimum error variance (MEV) estimate). The
standard deviation of the posterior probability function can also be
used to give an idea of the error in redshift resulting from the
photometric error (the $\sigma_f$ term in equation~\ref{likelihood} at
low signal to noise).

In likelihood-based methods, it is common to use a minimum error
variance (MEV) estimator (the mean redshift from the likelihood
function) rather than a maximum likelihood (ML) estimator.  This
estimator minimises the average deviation of photo-z values from the
true redshifts \citep{CADIS}. We argue here that in low
signal-to-noise situations it is best to avoid this step and adjust
one's analysis to work with $p(z)$ rather than any single estimation
for an object. In the high signal-to-noise situation the resulting
$p(z)$ can still be used to obtain such estimates if desired.

\subsubsection{Constructing a prior probability distribution}
\label{constructing_prior}
To make use of the Bayesian approach we need a method of calculating a
prior probability distribution $\mathcal{P}$
which incorporates our knowledge of the expected redshift
distribution at a given apparent magnitude.

Our prior probability is precalculated for each combination of a range
of apparent magnitudes (tailored to fully cover the range expected for
galaxies in a survey), for each SED and for each redshift. At every
$(z,S,R)$ location (of 177 points in $z$, 360 in $S$ and 50 in $R$)
the absolute magnitude $M$ necessary to result in an observation at
$R$ given a $\Lambda$CDM cosmology ($\Omega_M=0.3$,
$\Omega_\Lambda=0.7$) is calculated, using K-corrections derived from
the same galaxy templates used to determine photo-z colours. The three
variables $M$, $z$ and $S$ are used to derive the value of the galaxy
luminosity function (GLF) and this is converted to a density per unit
solid angle on the sky and per redshift interval (given the
$\log(1+z)$ spacing of the redshift points), again for a $\Lambda$CDM
cosmology. Allowing for the fact that $R$ is the corresponding
magnitude for the model flux $\hat{f}_R$, the resulting prior function
$\mathcal{P}(z|S,R)$ found from this table can be directly used in
equation~\ref{likelihood} .

The COMBO-17 GLFs are calculated at redshift bins centred on $z=0.3,
0.5, 0.7, 0.9$ and $1.1$. Photometric redshifts (utilising all 17
bands) and derived K-corrections are used to determine these GLFs. The
functions used in the prior are calculated for arbitrary redshift by
taking parameters $(M^*,\phi^*, \alpha)$ fitted by simple functions to
the five COMBO-17 values, accounting for their respective error bounds
(except in the case of $\alpha$ which was held constant in redshift
but not colour type when the luminosity functions were originally
fitted to the COMBO-17 data). These functions are a second order
polynomial in $\log \phi^*$ and a linear fit in $M^*$, chosen for a
low-order but good fit through the points.  In the case of $\phi^*$
the fit to the logarithm was chosen to prevent the galaxy number
density from taking on non-physical negative values. Furthermore the
evolution of $M^*$ is halted at redshifts greater than 1.0. This
constraint on the evolution is consistent with~\citet{Cohen} who gives
$\log L^*(R) = 37.37 \pm 0.25$ at $0.8<z<1.05$ and $\log L^*(R)=37.30
\pm 0.40$ at $1.05<z<1.50$. \citet{Ilbert} also find no evidence of
$M_R^*$ evolving beyond $z=1$, except for some slight brightening in
the case when $\alpha$ is fixed, with most evolution occurring in
bluer bands.

The galaxies in each type are assumed to be spread evenly within the
SED range covered by the original colour cuts separating each
population when converting between the individual 360-point SED
classification scheme of the templates and the summed 3-population
classification of the GLFs.

Whilst a prior constructed in this manner is technically
cosmology-dependent, being founded on a $\Lambda$CDM luminosity
distance, this is the same cosmology originally used to determine the
GLFs from COMBO-17, and therefore much of the dependence on the
cosmological model cancels out. This leaves only the potential for
small differences due to fits to the parameter evolution and is not a
concern compared to the uncertainty already present in the GLFs.

\subsection{Comparison of likelihood and Bayesian methods}
\subsubsection{Statistical tests}
We examine the effectiveness of the photometric redshift estimation in
three ways. We define \textit{sensitivity} as the fraction of galaxies
in some spectroscopically-measured redshift range that that have
photometric redshift correct within some tolerance.  However, we are
also interested in the fraction of galaxies at some {\em estimated}
redshift that are correct within a specified tolerance.  Following the
terminology used in clinical trials, we call this statistic the
\textit{specificity}.

Finally, we have already discussed the possibility that estimations
may be biased, and we examine this possibility by considering the mean
expected photometric redshift for samples binned by spectroscopic
redshift and how the posterior probabilities vary with spectroscopic
redshift. We also compare redshift distributions for samples of
galaxies in various magnitude ranges in Section~\ref{section_faint}.

\subsubsection{Comparison at bright magnitudes: MEV estimation}
At sufficiently bright magnitudes, $R\la 23$, we may compare directly
photometric redshift estimates with spectroscopically measured redshifts
from VVDS.  This comparison has the great strength of being based on 
a known ``ground truth'', but unfortunately does not probe to the
faintest magnitudes that are of interest in weak lensing studies.

\begin{figure}
\resizebox{0.8\linewidth}{!}{\includegraphics{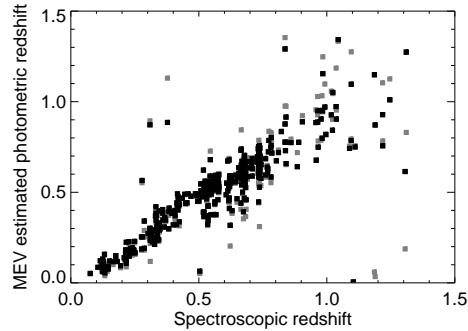}}
\caption{Minimum error variance (MEV) photometric redshift plotted against VVDS spectroscopic redshifts. Grey points represent those estimated with no magnitude prior, black points those estimated with the application of a magnitude prior. Limited to $R<23$.}
\label{UBVRI}
\end{figure}

Fig.~\ref{UBVRI} shows MEV photometric redshift estimates based on
$UBVRI$ data where $R<23$ for both a standard template technique and with a Bayesian
prior applied. The standard template technique used is an MEV estimate equivalent to using an uninformative flat prior so that
\begin{equation}
\mathcal{L} \propto \int^{\infty}_0 \exp \left[  -\frac{1}{2} \left(
  \frac{\mathbf{f}-\hat{f}_R
    \boldsymbol{\hat{\lambda}}}{\boldsymbol{\sigma_f}} \right) ^2
  \right] d\hat{f}_R
\end{equation}
where we again marginalise over the model flux $\hat{f}_R$.

For this set of brighter objects (60 per cent of the sample), redshift
estimates are largely unchanged. Some outlier objects are have
improved redshifts, but equally some are scattered further from the
true value. The overall standard deviation of
$z_\mathrm{photo}-z_\mathrm{spec}$ for each technique is not
substantially changed, being $27$ per cent worse for the non-Bayesian case,
but this is predominantly due to changes in outlier estimations.

\begin{figure}
\resizebox{0.8\linewidth}{!}{\includegraphics{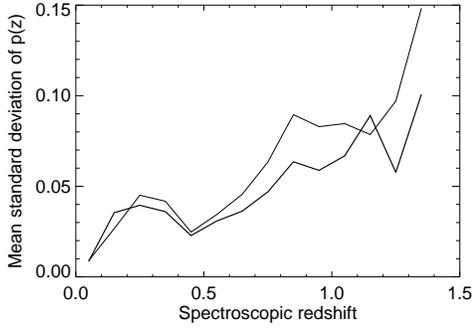}}
\caption{Mean $\sigma_z / (1+z)$ (standard deviation of $p(z)$) in the case of no prior (upper line) and in the case of a magnitude prior (lower bold line). Results are binned in redshift intervals of 0.1 from z=0 to z=1.4 for objects where $R<24.5$}
\label{sigmaz_reduction}
\end{figure}

\begin{figure}
\resizebox{0.8\linewidth}{!}{\includegraphics{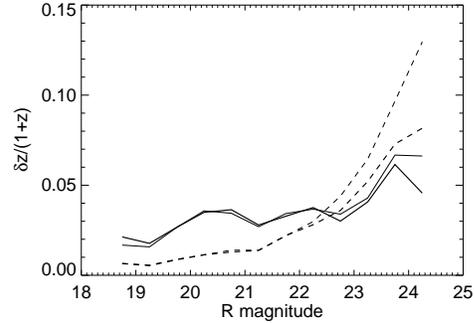}
}
\caption{RMS redshift error as $\delta z / (1+z)$ as a function of R
  magnitude without prior (thin) and with (thick) along with width of
  probability functions scaled by $1/(1+z)$ (dashed lines), binned in
  0.5 magnitude intervals from $R=18.5$ to $R=24.5$,
  $0<z<1.4$. Outliers, defined as objects more than 3s.d. (calculated
  iteratively) from the mean $\delta z / (1+z)$, are excluded (6 and 8
  per cent for the Bayesian sample and non-Bayesian sample respectively).}
\label{ubvri_magnitude_redshifterrors}
\end{figure}

As signal-to-noise decreases there is a substantial increase in the
advantage gained by applying a prior, as the width of the $p(z)$
function decreases (Fig.~\ref{sigmaz_reduction}). This is more
substantial at high redshift since under the standard likelihood
approach fainter objects (which dominate at high redshift) will
produce a broader $\mathcal{L}(z)$, and it is in this region where the
Bayesian approach provides the majority of the information. This
effect is seen more clearly when data are binned by $R$ magnitude in
Fig.~\ref{ubvri_magnitude_redshifterrors} where we also see the actual
$\delta z/(1+z)$ of MEV estimates.  The relatively low $\delta
z/(1+z)$ errors at the faintest magnitudes may be due to the
high-confidence spectra from VVDS at such faint magnitudes being
dominated by objects with particularly high-contrast spectral features
suitable for making unusually good photometric redshift estimates, and
should not necessarily be relied upon as being representative of the
errors for faint objects in general.

For these MEV estimates we have removed redshift offsets of
$0.028\times(1+z)$ and $0.023\times(1+z)$ for the non-Bayesian and
Bayesian sets respectively. Such offsets are a result of the
systematic effects described below and are dependent on various
aspects of the photometry calibration. They can not be relied upon to
be field independent and would in practice require some spectroscopic
calibration to correctly remove. 

In addition, we have eliminated outliers where $z/(1+z)$ has been
misestimated by more than 3 standard deviations, calculated
iteratively in unit magnitude bins from 18.5 to 24.5 (eliminating also
two objects beyond 24.5). 51 outliers are eliminated in the
non-Bayesian approach and 37 in the Bayesian approach.  The error
estimated from the width of the $p(z)$ distribution underestimates the
true redshift error at high signal to noise, where the largest sources
of errors are systematic effects such as relative calibrations of the
passbands and variations in object spectra that are not represented in
the templates. At lower signal to noise the range of possible colours
covers the templates well enough that template incompleteness and
passband uncertainties are less significant and allow for more
accurate estimations of error based on the $p(z)$ width alone (see
also \citealt{COMBO-17} for a more detailed discussion of the
properties of such estimation errors). In the case of the sample used
here this transition to noise-dominated errors occurs at $R \lesssim
23$ and is the reason the $R<23$ cut was chosen in the examination of
MEV estimates earlier.

\begin{figure*}
\centering{
  \subfigure[ ]{\includegraphics[width=0.32\linewidth]{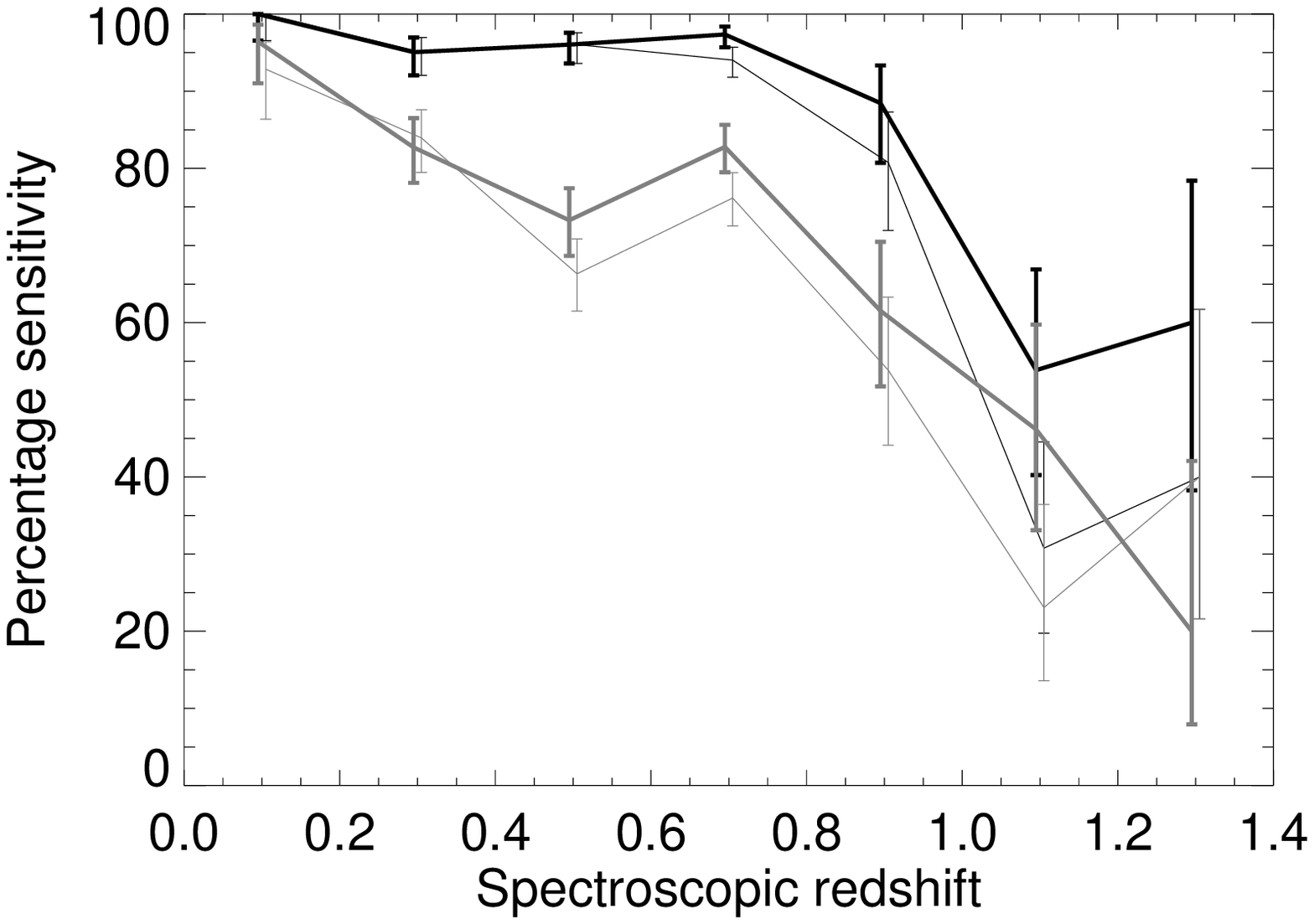}} \hfill
  \subfigure[ ]{\includegraphics[width=0.32\linewidth]{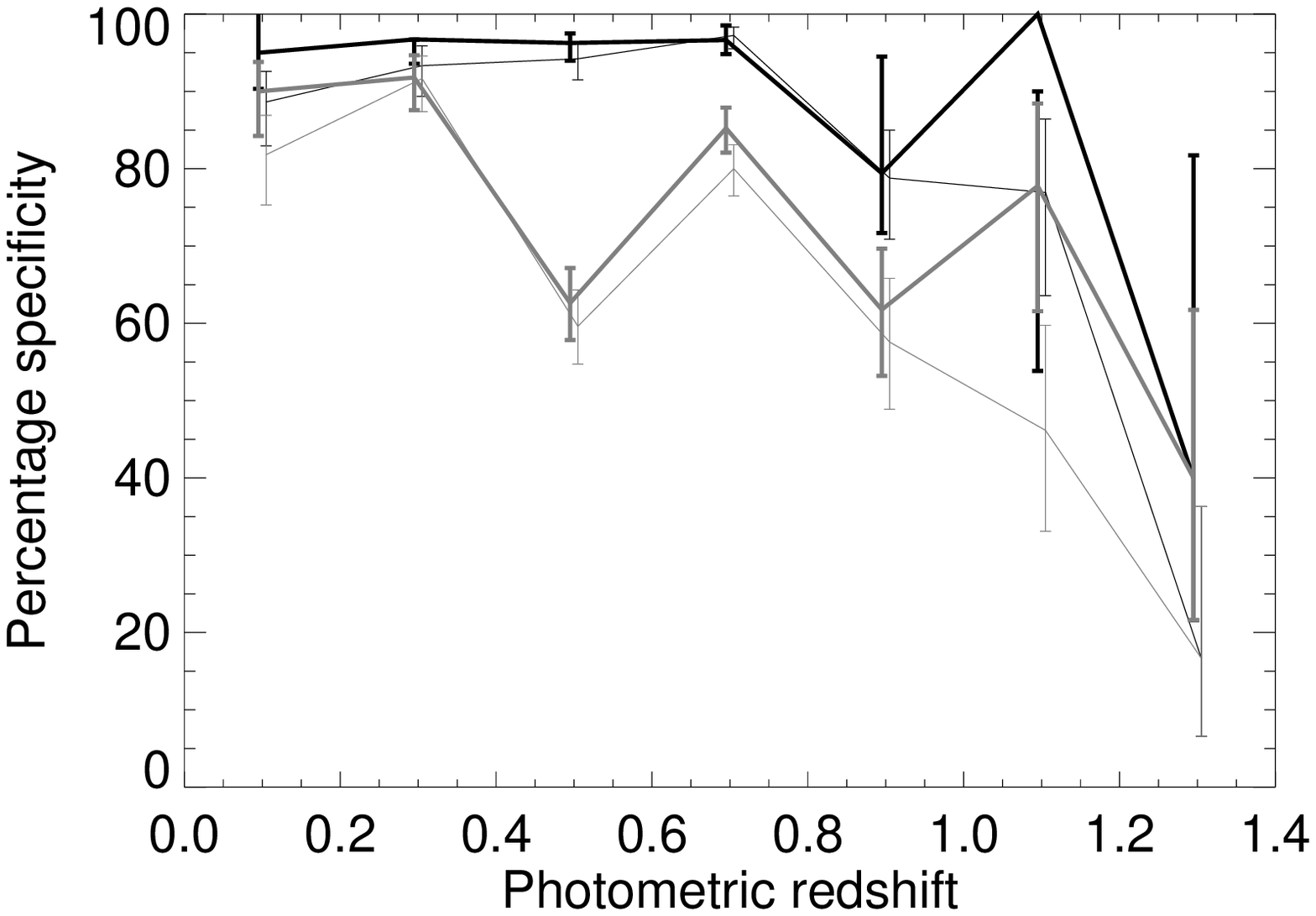}} \hfill
  \subfigure[ ]{\includegraphics[width=0.32\linewidth]{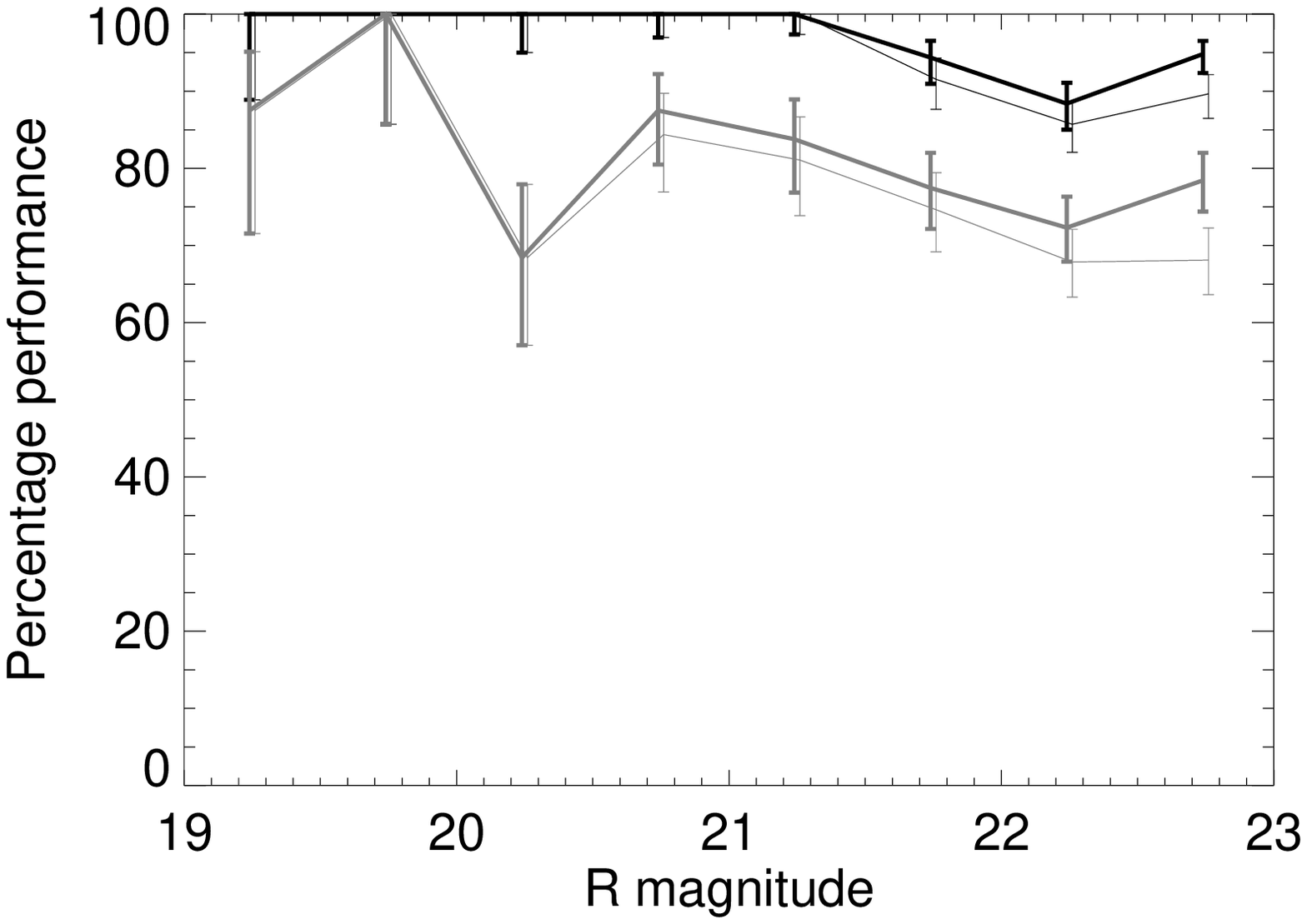}} }
\caption{(a) Sensitivity, (b) specificity and (c) performance by magnitude
of MEV photometric redshifts for $R<23$. Upper black lines 
show tolerance of $0.1\times(1+z)$ and lower grey lines show tolerance of
$0.05\times(1+z)$. In each case the thin line represents results
before a prior is applied and the thick line those afterwards. Error
bars are estimates under the assumption that objects fall in or out of
the tolerance according to the binomial distribution. In the case of performance by magnitude
the concept of sensitivity and specificity do not apply as they are a result of binning by either
photometric or spectroscopic redshift.}
\label{senspecperf_MEV}
\end{figure*}

Sensitivity and specificity for $UBVRI$ MEV estimates with and without
a prior is given in Fig.~\ref{senspecperf_MEV} for the 400 objects
with $R<23$. Outliers are included again, and we use a varying
tolerance proportional to $1+z$ (as a fixed resolution $\delta \lambda
/ \lambda$ translates to fixed $\delta z / (1+z)$). Overall 73 per
cent of objects have redshifts correct to $0.05\times(1+z)$ in the
non-Bayesian case and 78 per cent in the Bayesian case. 92 per cent
have redshifts correct to $0.1\times(1+z)$ in the non-Bayesian case,
and 94 per cent in the Bayesian case, showing that single-valued
estimates of redshifts are largely unchanged at brighter
magnitudes. Error bars are estimates based on the assumption that
objects fall in or out of the tolerance range according to the
binomial distribution and are shown in order to give a better idea on
the constraint of the true tolerance and can not indicate whether the
techniques are significantly different. Performance by magnitude is
also shown in Fig.~\ref{senspecperf_MEV} again showing broadly similar
performances between the two methods for MEV estimates at brighter
magnitudes, showing that gains are starting to be made as we approach
the point at which these estimates will start to break down.

\subsubsection{Using the full posterior probability distribution}
\label{usingfullprior}
At low signal-to-noise it becomes appropriate to avoid single-valued
estimations (see Section~\ref{section_faint}). To compare sensitivity
and specificity in the case where we aim to use summed normalised
posterior distributions we define each as the percentage of the
distribution within the given range, so sensitivity is defined as
\begin{equation}
\frac{\sum_i \int P_i(z) W_i(z) dz}{N_i}
\end{equation}
where $i$ is the set of objects in a given bin of spectroscopic redshift, and specificity is defined as
\begin{equation}
\frac{\sum_i \int_{z_A}^{z_B} P_i(z) W_i(z) dz}{\sum_i \int_{z_A}^{z_B} P_i(z) dz}
\end{equation}
where $i$ is the set of all objects, $z_A$ and $z_B$ denote the limits
of the bin in consideration and we use a window function $W_i(z)$ as
an additional constraint on the limits of the integrals so that
\begin{alignat*}{1}
W_i(z) &= 1 \text{ where } |z-z_{\mathrm{spec}_i}|<T(1+z_{\mathrm{spec}_i}) \\
       &= 0 \text{ at all other points}
\end{alignat*}
for a given tolerance $T$. With these definitions sensitivity and
specificity are sums over equivalent regions in $z_\mathrm{spec}$ and
$z$ to our earlier definitions.

Furthermore we define a performance binned by magnitude as
\begin{equation}
\frac{\sum_i \int P_i(z) W_i(z) dz}{N_i}
\end{equation}
where $i$ is the set of objects in a given magnitude bin.

\begin{figure*}
\centering{
\subfigure[ ]{\includegraphics[width=0.32\linewidth]{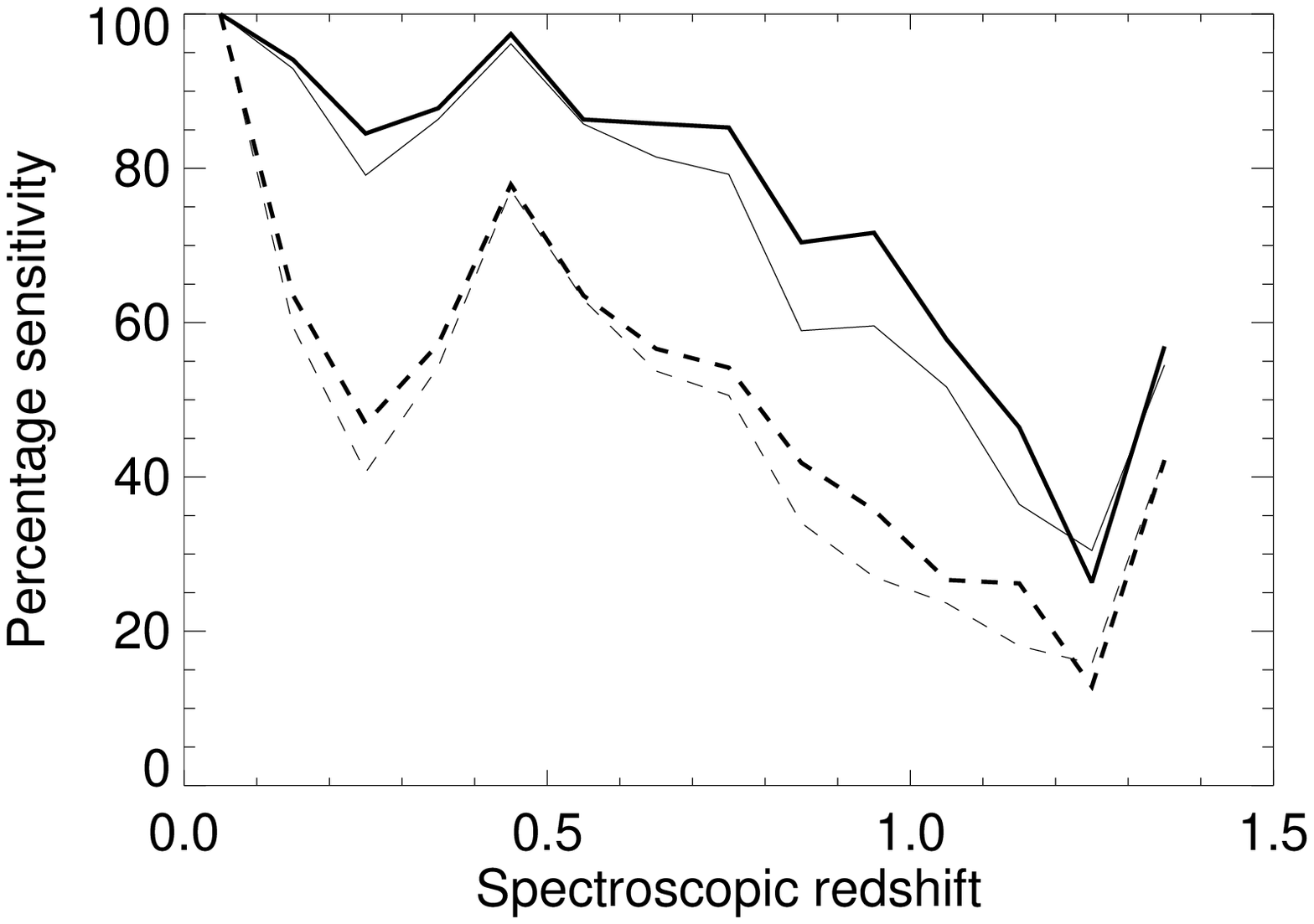}}\hfill 
\subfigure[ ]{\includegraphics[width=0.32\linewidth]{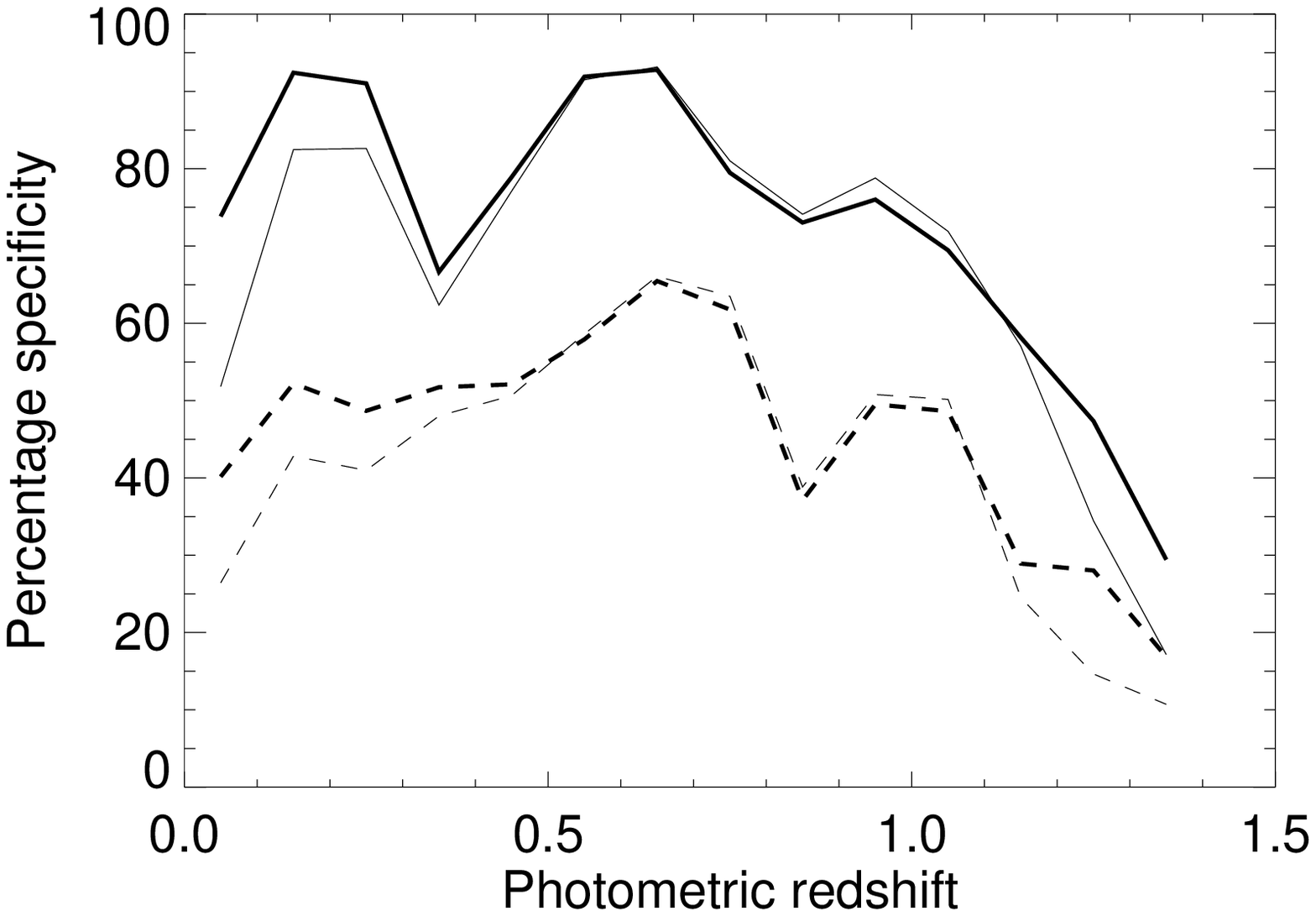}}\hfill
\subfigure[ ]{\includegraphics[width=0.32\linewidth]{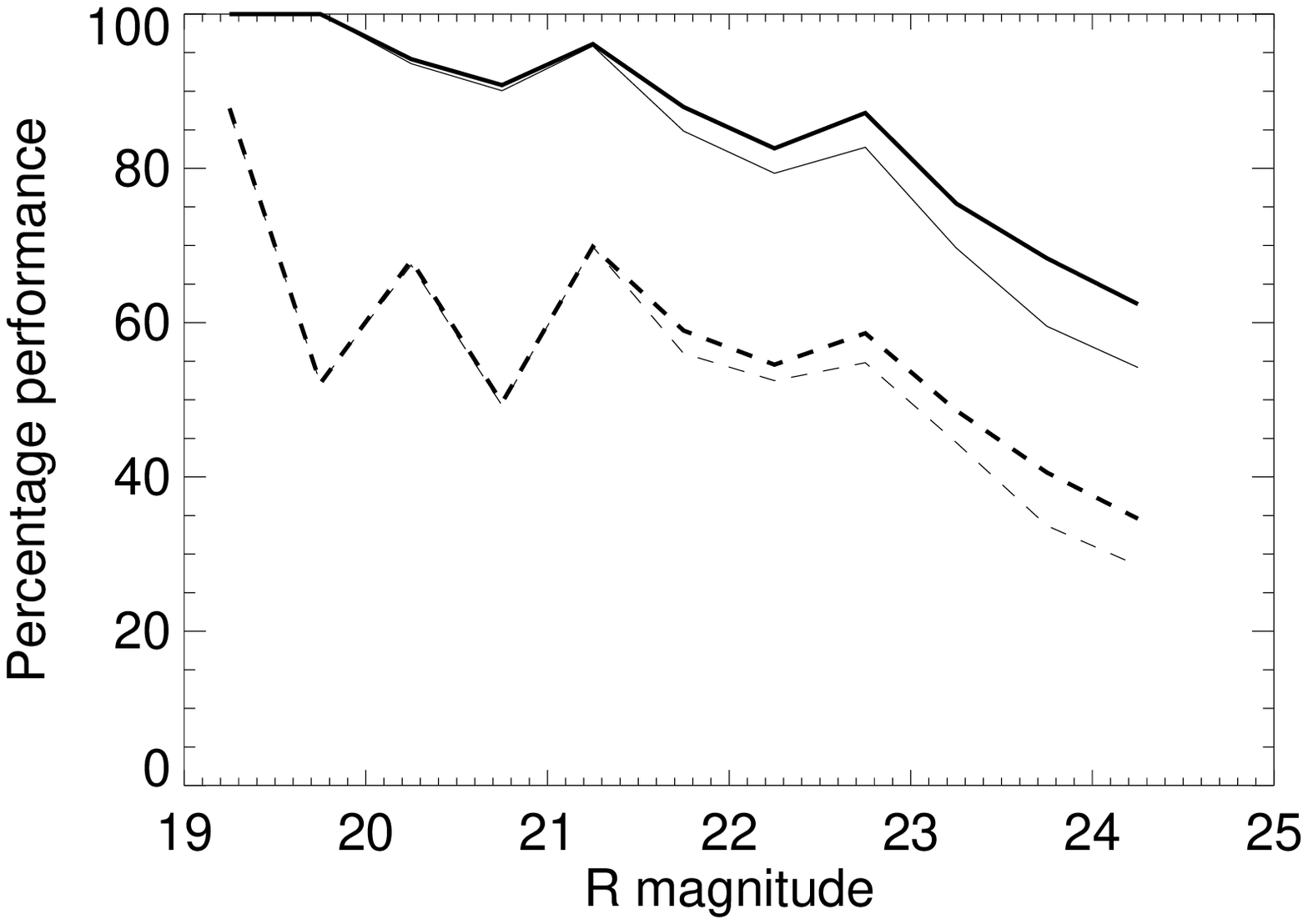}}
}
\caption{(a) Sensitivity, (b) specificity and (c) performance of
  photometric redshifts, full posterior definition, $R<24.5$. Upper
  solid lines show the percentage of the $L(z)$ or $p(z)$ function
  within a tolerance of $0.1\times(1+z)$ and lower dashed lines show
  tolerance of $0.05\times(1+z)$. In each case the thin line
  represents results before a prior is applied and the thick line
  those afterwards.}
\label{senspecperf_full}
\end{figure*}

Sensitivity, specificity and performance plots in this case are given in
Fig.~\ref{senspecperf_full}. Sensitivity
shows a small gain at almost all redshifts, and differences in
specificity are also generally small. Overall performance is 49 per cent
within a $0.05 \times (1+z)$ tolerance without a prior and 53 per cent
with, and 76 per cent within a $0.1 \times (1+z)$ tolerance without a prior
and 81 per cent with.

\begin{figure}
\resizebox{0.8\linewidth}{!}{\includegraphics{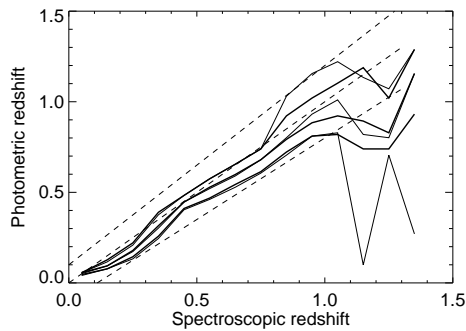}}
\caption{Plots of the 25th, 50th and 75th percentile positions of
summed posterior probabilities in redshift bins of 0.1 for $R<24.5$ in
the case of no prior (thin lines) and a Bayesian prior
(thick lines). Dashed lines indicate a line of
$z_\mathrm{photo}=z_\mathrm{spec}$ and a region $\pm 0.1 \times (1+z)$
tolerance.}
\label{fullpz}
\end{figure}

Fig.~\ref{fullpz} show the 25th, 50th and 75th percentiles of summed
$p(z)$ distributions in the cases without a prior and with a Bayesian
prior over a range of spectroscopic redshifts and gives a clearer
indication of regions where the prior has most effect. At low
redshifts where objects tend to be brighter there is little difference
but at higher redshifts and for fainter objects the estimates become
much broader and generally occupy a range from about $z\approx 0.7$
upwards. The Bayesian estimates help resolve a bimodality with much
lower redshifts around this region (which is the cause of the 25th
percentile of the non-Bayesian distribution lying at such low
redshift) and give a higher proportion of the estimation to the higher
redshifts, although neither approach gives a particularly good match
between the median distribution redshift and the spectroscopic
redshifts.

\begin{figure*}
\centering{
\subfigure[ ]{\includegraphics[width=0.32\linewidth]{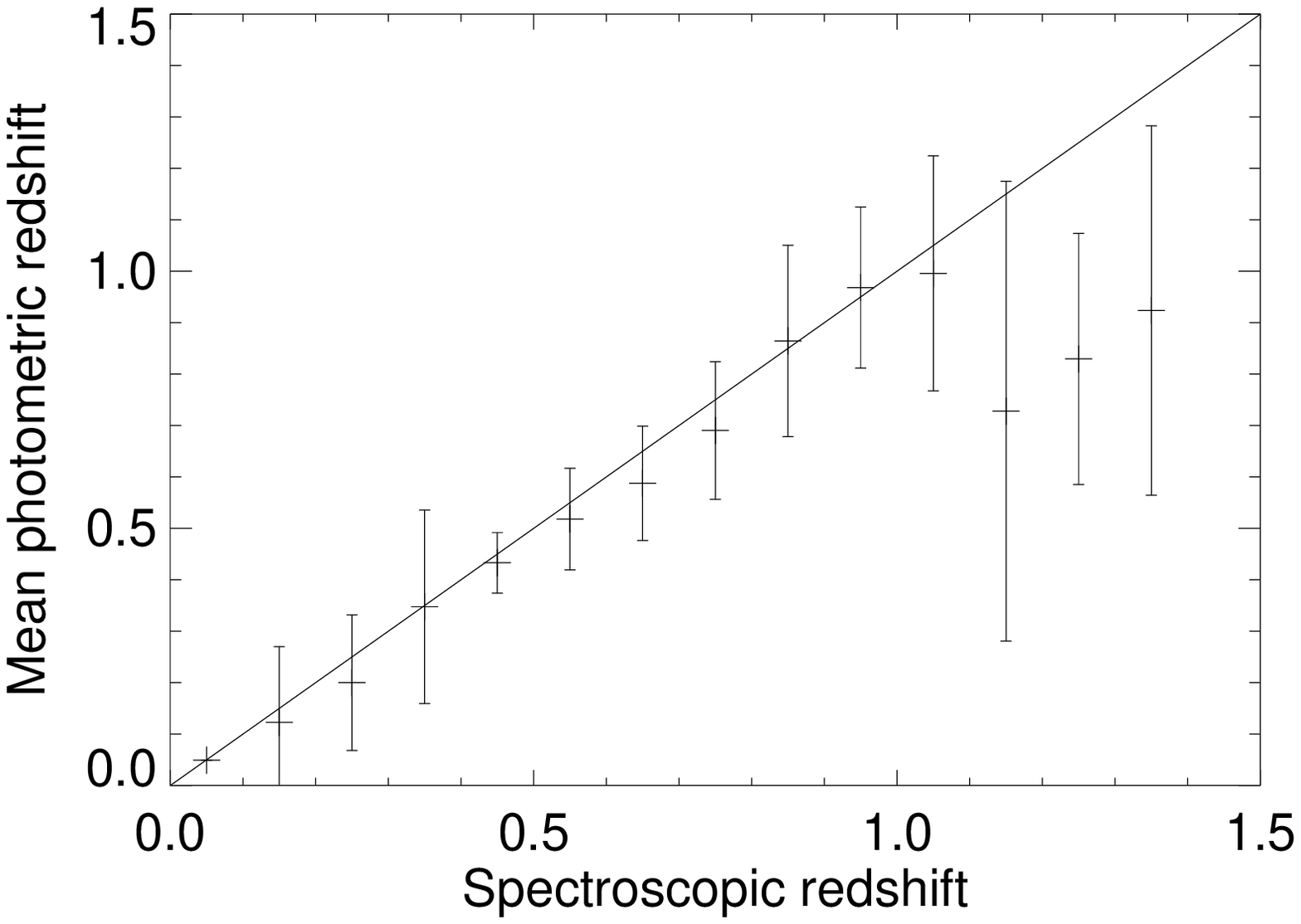}} \hfill
\subfigure[ ]{\includegraphics[width=0.32\linewidth]{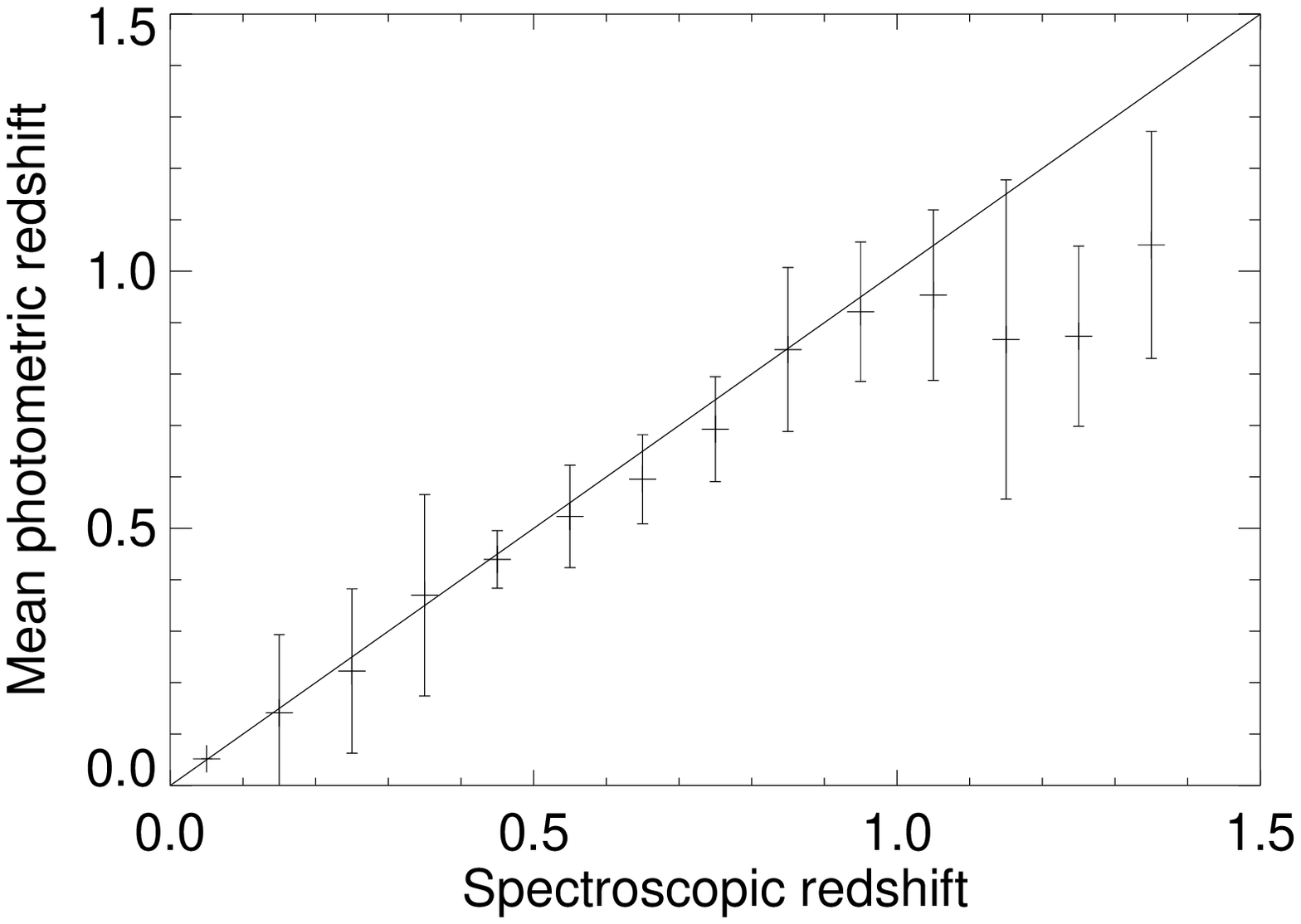}} \hfill
\subfigure[ ]{\includegraphics[width=0.32\linewidth]{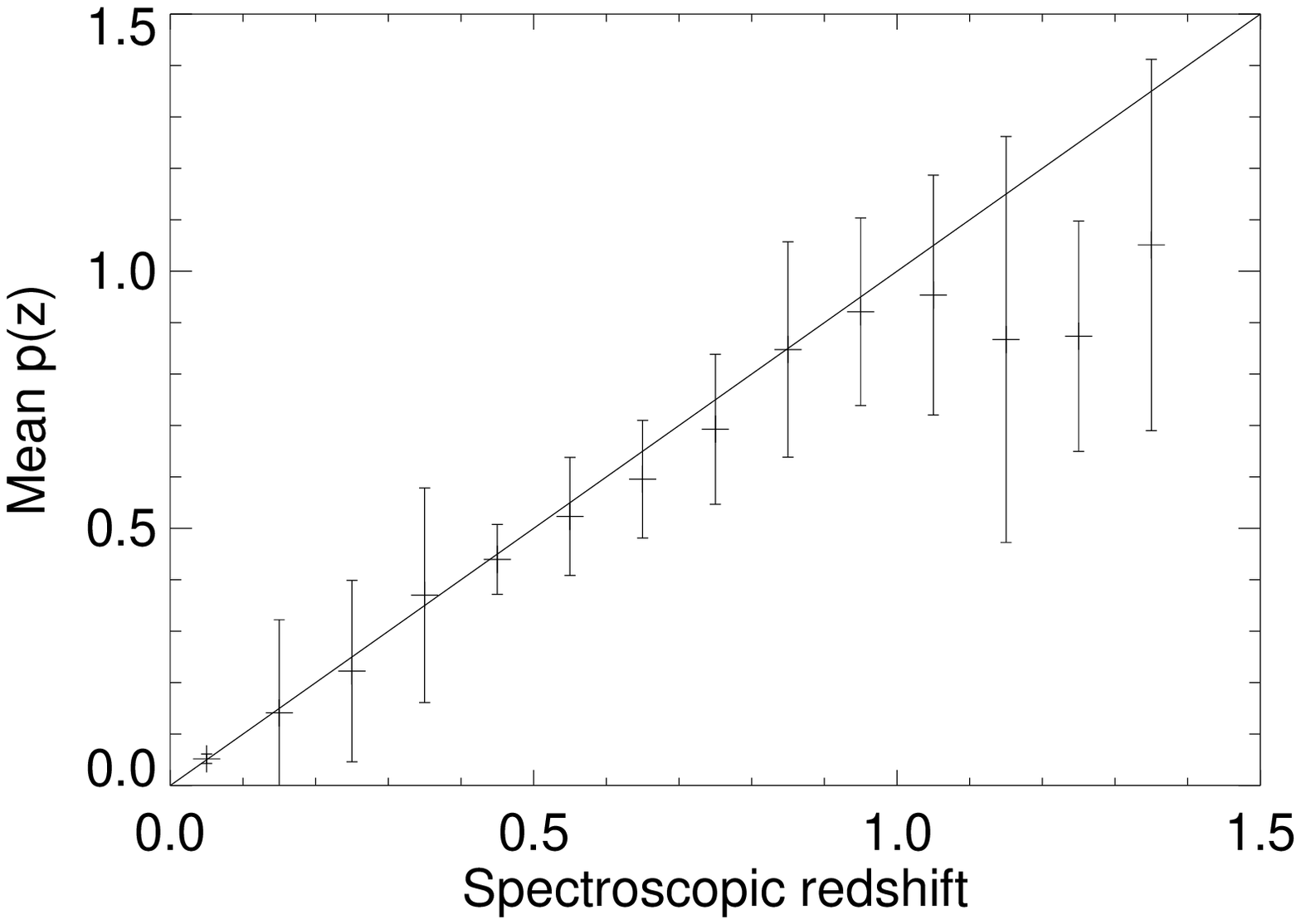}}}
\caption{Mean estimated redshift for $R<24.5$ binned by spectroscopic
redshift (a) without prior, (b) and (c) with prior. Error bars
represent the $1\sigma$ spread in MEV estimates for (a) and (b) and
the mean width of the summed $p(z)$ for (c). Lines indicate
$z=z_\mathrm{spec}$ where there would be no bias.}
\label{meanz}
\end{figure*}

As well as reproducing the distribution as accurately as possible, we
must also examine whether selections from this distribution are
unbiased representations of the true underlying redshifts. Clearly
from examining Fig.~\ref{fullpz} there is some degree of bias
present. To check the severity of the bias, we also examine the mean
photometric redshift in the two MEV cases and the mean of the summed
posterior probabilities and the width of this distribution in the
Bayesian case also in spectroscopic redshift bins of width 0.1
(Fig.~\ref{meanz}). We perform a least-squares fit to determine a bias
$b$ where $\langle 1+z_\mathrm{photo} \rangle = b
(1+z_\mathrm{spec})$, weighting by the number of objects in each bin
and inversely to the variance of MEV estimates or the width of the
mean $p(z)$ estimate squared. Since the ability to make a reasonably
unbiased estimate appears to break down above $z\approx1$ as noise
increases and the $z<0.1$ contains too few objects to rely upon we
consider $0.1<z<1.0$ only. Non-Bayesian MEV estimates give a fit of
$b=0.976 \pm 0.003$, Bayesian MEV estimates give $b=0.977 \pm 0.003$
and summed posterior probabilities give $b=0.978 \pm 0.003$,
indicating that there is no significant change in bias between the
three approaches in that redshift range.

As well as the bias in each bin, the spread in redshift estimates is
also a concern. In Fig.~\ref{fullpz} in the Bayesian case, half the
probability distribution is contained within a region of $\delta z
\approx 0.05$ at $0.1<z<0.2$ rising to $0.21$ at $0.9<z<1.0$. At
higher redshifts the width is less stable, ranging from 0.28 to 0.45
at $z>1.0$. In comparison, the non-Bayesian case has $\delta z \approx
0.04$ at $0.1<z<0.2$ rising to $0.35$ at $0.9<z<1.0$, and from there
as high as $1.03$ at higher redshifts.

A uniform distribution of redshifts in these bins would have half the
distribution in $\delta z = 0.05$, so there is a spread as would be
expected due to the errors arising from estimating photometric
redshifts, which presents a concern. In addition, there are
significant outlier peaks in the probability distribution (some of
which are picked up in Fig.~\ref{fullpz} by the 25th percentile of the
non-Bayesian distribution) which could disrupt an analysis, but the
impact these regions have is reduced in the Bayesian approach, and
this advantage is strongest for the faint galaxies where traditional
methods break down.

Fig.~\ref{meanz} shows this also for MEV estimates, although the error
bars here are influenced by the outlier regions mentioned above. In
particular, the low redshift end shows larger error bars than might be
expected from Fig.~\ref{fullpz} as the second moment of the
distribution is more sensitive to these outliers than the percentile
positions. The MEV estimates also show less spread than the full
posterior approach due to undersampling of the tails, and whilst this
might appear initially a useful property we show in
Section~\ref{section_faint} that this can be highly problematic as
signal-to-noise deteriorates. We return to this issue, and that of
bias and higher order issues of the distribution in weak lensing
analyses in Section~\ref{lensingcalculations}.

\subsection{Effects on redshift estimates from errors in the prior}
\label{priorerrors}
There is a concern that the prior may not be the correct choice due to
both measurement errors in the luminosity functions used in its
calculation, and also due to systematic effects that might arise from
the choice of fitted models to the luminosity functions and the
treatment of the three types of galaxies used in the luminosity
function calculation (such as the choice of spreading the luminosity
function density evenly across the SED ranges).

A full analysis of the effect of measurement errors on the prior is
computationally infeasible, and similarly attempting to make a full
assessment of the impact of any effects of modelling choices and
systematic effects arising from the method used to assign the
luminosity function across SED types would be similarly
infeasible. However, some assessment of the errors that would arise
can be made by investigating a few special cases.

We have reanalysed the above sections with four further priors. Two of
these have a faint end slope $\alpha$ of the red galaxies adjusted by
$\pm 1 \sigma$, and two have the $z=1.1$ value of $\phi^\star$ of the
red galaxies adjusted by $\pm 1 \sigma$. We adjust these parameters in
preference to any others as the parameters most likely to make large
changes to the prior. 

High redshift values are chosen since adjustments to the low redshift
parameters will have less effect, since most objects will be expected
at higher redshift except for brighter objects which will already have
well-constrained estimates and which will therefore be insensitive to
the form of the prior.

Also, an overall shift in $\phi^\star$ affects the normalisation of
the prior only, so we vary the value of this parameter for one class
of galaxies rather than for all.

We ignore the variation in $M^\star$ typically has a lower impact on the
resulting $\phi$ than the value of $\phi^\star$ does, and also since
evolution of this parameter is halted at $z=1.0$ the models are
more limited in the effect that variations in the $z=1.1$ point may
have.

These four priors produce very little difference in the overall
performance of the redshift estimates, with changes on the order of
tenths of a percent corresponding to just a few objects falling in or
out of the $0.05\times (1+z)$ and $0.1\times (1+z)$ tolerance
bounds. For MEV estimates all priors have $78\%$ within $0.05\times
(1+z)$ and $94\%$ within $0.1\times (1+z)$ for all but the prior with
$\alpha$ raised, which performs marginally better at $95\%$
($R<23$).With the full priors the figures are $53\%$ for all priors at
$0.05\times (1+z)$ tolerances and $81\%$ for the original and
$\alpha-1\sigma$ priors and marginally better with $82\%$ for the
$\alpha+1\sigma$ and $\phi^\star(z=1.1)\pm 1\sigma$ priors ($R<24.5$).

This lack of significant difference can be partly ascribed to the fact
that these are all brighter objects, and we examine this issue again
in section~\ref{lensingcalculations} when we use simulations as deep
as $R<25.5$ when this issue can be expected to have a greater
impact. These results also of course only vary one parameter by $1
\sigma$, and in reality some noise will apply to every parameter, but
the small changes from varying these parameters which were chosen for
maximum effect would not be expected to rise to problematic levels,
and future constraints on luminosity function parameters can be
incorporated to reduce the impact of the lack of knowledge we have of
the form of the true underlying distribution of galaxies.

\section{Photometric redshifts with high noise photometry}
\label{section_faint}
\begin{figure*}
\subfigure[ ]{\includegraphics[width=0.48\linewidth]{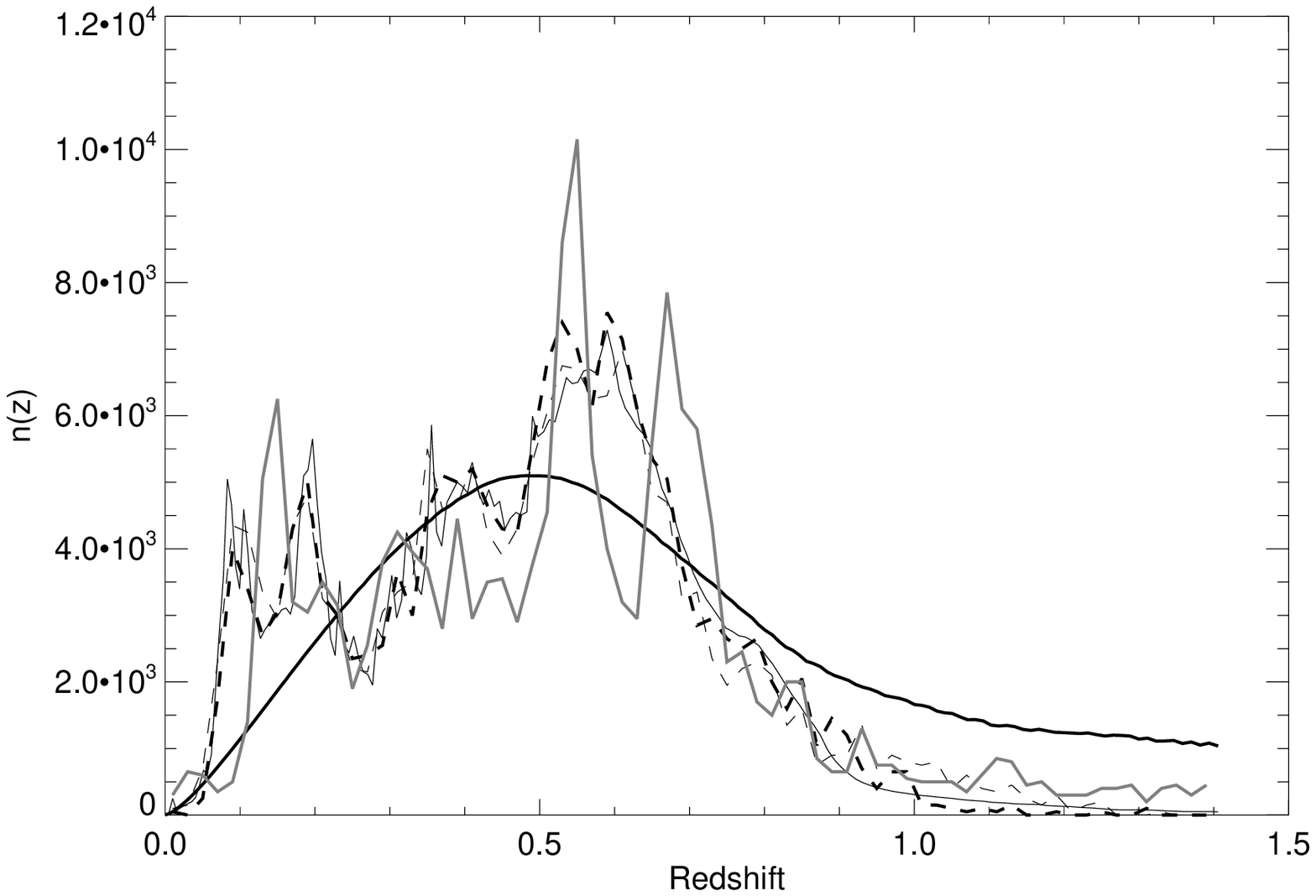}}
\subfigure[ ]{\includegraphics[width=0.48\linewidth]{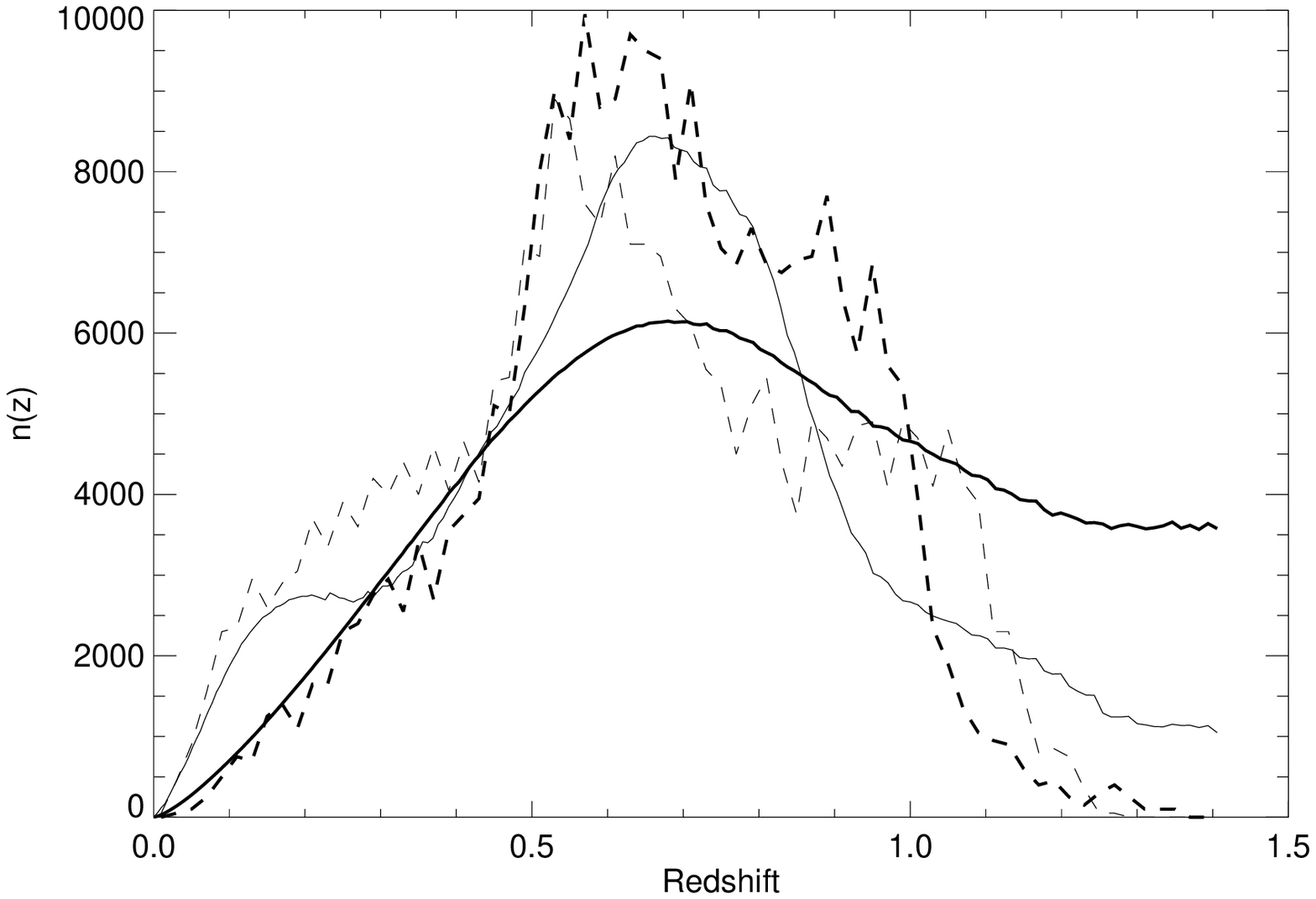}}
\subfigure[ ]{\includegraphics[width=0.48\linewidth]{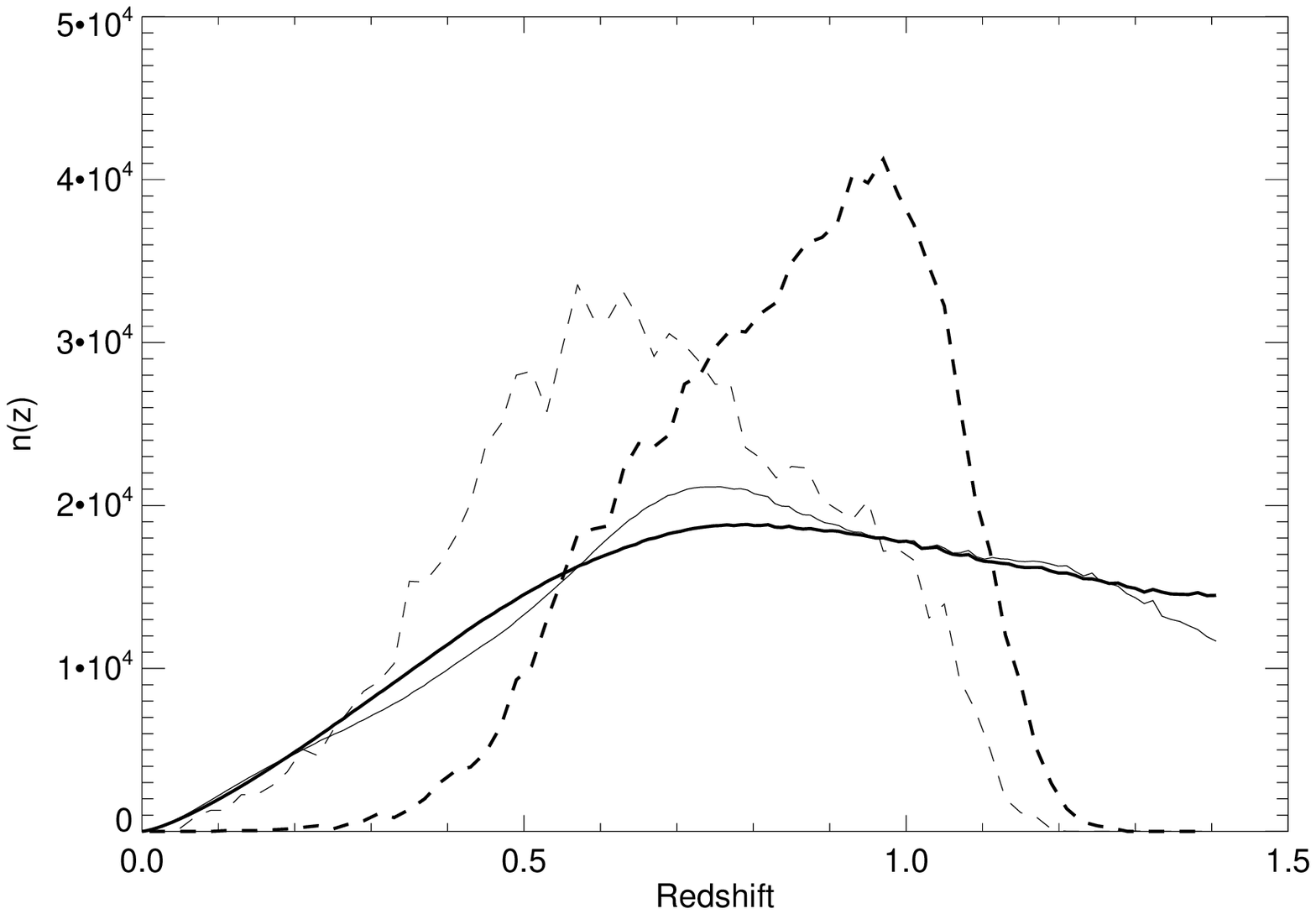}}
\caption{Estimated $n(z)$ distributions from a non-Bayesian MEV
estimation (dashed line), Bayesian MEV (thick dashed line), summed
posterior probability approach (thin continuous line) and a pure prior
based solely on knowledge of the magnitude selection (thick continuous
line). The distributions are for (a) $17<R<23$ (3,390
objects), (b) $23<R<24$ (5,285 objects) and (c) $24<R<25.5$ (18,885
objects). The fluctuations in the prior and summed posterior
probability are slight artifacts from the process of calculating the
prior distribution. For the $17<R<23$ set COMBO-17 MEV estimates
are also included, shown as a solid grey line. 3 per cent of the sample do
not receive a COMBO-17 estimate, and for the fainter samples this
proportion rises considerably, so we do not show the distributions in
these cases. All three sets of MEV estimates are binned in redshift
intervals of $0.02$.}
\label{distributions}
\end{figure*}

Perhaps the most significant benefit of a full posterior approach is
in the case where redshift estimates traditionally break down owing to
photometric noise and $L(z)$ becoming increasingly flat. In this
situation one might conventionally set aside all objects with such
poor photometry and in a weak lensing analysis simply estimate a
single redshift distribution for all such objects. As faint objects
are both more numerous and are generally at higher redshifts than more
luminous ones the need to deal with them in an optimal manner is of
greater concern than attempting to improve the already reasonably
well-constrained redshifts of brighter objects.

The VVDS sample does not provide enough objects at a faint enough
magnitude to study this effect so instead we consider a sample of
28,597 objects from the CDFS field of COMBO-17 with $17\le R<25.5$ and
is classed by COMBO-17 as a galaxy based on photometry
(\citealt{COMBO-17} describes this in more detail). Some objects have
photometry that does not fit the templates sufficiently well and have
extremely low likelihoods that produce a $p(z)$ distribution that due
to computer rounding is zero at all redshifts, and these are
excluded. There are not in general spectroscopic redshifts available
for this sample, and for fainter objects reliable photometric
redshifts are not available either (even when using all 17
bands). However, we can consider the distribution of photometric
redshifts and compare them to expectations. Plots of the distributions
of $UBVRI$-based MEV estimates along with the summed posterior
probabilities and an expected distribution derived from the prior in
three magnitude ranges are shown in Fig.~\ref{distributions}, along
with 17-band MEV estimates from COMBO-17 for the brightest set.

For brighter objects the estimates all converge on a result that
includes more details of the large-scale structure in the field
(\citealt{Gilli}, \citealt{VVDS} and \citealt{COMBO-17} describe the CDFS as
having rather pronounced large-scale structure). The peaks at $z
\approx 0.15$ and $z\approx 0.6$ are both seen in the distribution of
COMBO-17 photometric redshifts for example, although they are not so
well resolved in the broad-band only estimations. The adjustments made
by the prior are either small or, in the case of outlier correction,
too rare to have a large effect on the overall distribution.

In the faint regime of $R>24$ the MEV estimates are clearly distorted
by the inability to estimate a high redshift from a very broad $L(z)$
function (since $L(z)=0$ at $z>1.4$) and to a lesser extent the same
effect at low redshift, and as a result the estimates cluster at the
centre of the range of allowed estimates giving a distribution that is
too sharply peaked. In the non-Bayesian case the relative coincidence
of the peak of the distribution with that of the expected distribution
is not due to any fundamental link between the two but that the
expected distribution's peak happens to fall near the centre of the
estimation range. Even in the case of Bayesian MEV estimates the
$p(z)$ distribution is still broad enough that the estimates cluster
strongly in the central region of the estimation. These failures to
make a good estimate of the distribution make MEV estimations unusable
for faint objects. Maximum-likelihood (or
maximum-posterior-probability) estimators would suffer a similar
problem (although in the non-Bayesian case we would not expect the
distribution to be so sharply peaked, it would still have little, if any,
connection to the true distribution). Having a distribution that is
affected by this issue, whether as extremely as demonstrated in
Fig.~\ref{distributions} or to a lesser degree, would cause a bias in
estimates of cosmological parameters from weak lensing measurements.

In contrast, the summed posterior distribution does not lose the
contributions at the extremes and follows the expected distribution
but with some modulation from any information that can be obtained
from the photometry. Most importantly the severe bias in the
distribution is eliminated and whilst there may be some present from
the prior itself being generally biased compared to the true
distribution, this could be reduced by constructing new priors as
information on luminosity functions improves. We have therefore in the
summed posterior probability approach moved smoothly from a
well-constrained photometric redshift distribution towards a broadly
estimated distribution providing a unified approach without having to
decide upon a limiting accuracy of photometric redshifts beyond which
we deal separately with any objects with poor photometry.

\section{Sample priors}
\label{sampleprior}
An alternative approach to dealing with this problem of bias is the
``sample prior'' described in Section~\ref{choiceofestimator} - a
choice of prior that does accurately produce a set of estimates that
reproduces $n(z)$ accurately, albeit at the cost of individual
redshift estimates. This sample prior would have to be constructed
through Monte Carlo simulations. By using the Bayesian prior to
produce a sample set of simulated objects with a suitable overall
distribution with colours and magnitudes based on the photometric
redshift templates, and scattering these colours by an amount that
might be expected from typical photometric errors we can construct a
large simulation of a multicolour survey. These can be fed through a
standard non-Bayesian redshift estimation algorithm which will scatter
the original distribution to some estimated distribution and a
suitable prior can be constructed that would convert the resulting
estimated distribution back to the original.

This presents a practical problem in that the space to be sampled is
very large, but since we expect $n$ to vary smoothly with both $z$ and
$R$, and to a lesser extent SED we can still produce an attempt at a
sample prior through fitting polynomial functions to the distributions
to smooth over undersampled regions. An attempt at constructing a
sample prior in this way succeeds in producing a distribution estimate
approximately equivalent to a non-Bayesian approach but not as good as
the summed posterior approach. Whilst this might be improved upon by
constructing an improved prior the computational difficulty involved
would outweigh the marginal gains that might be expected over the
summed posterior estimation.

\section{Effect on lensing calculations}
\label{lensingcalculations}
To illustrate the effect of distortions in redshift distribution
estimations on weak lensing studies we consider the predicted value of
the critical surface mass density $\Sigma_\mathrm{crit}$ for lensing, which
provides a scale by which the dimensionless mass surface density
$\kappa$ can be converted to an estimated surface density. For a
particular combination of lens and source redshifts, this
critical density is given by
\begin{equation}
\Sigma_{\mathrm{crit}} = \frac{c^2}{4 \pi G D_d} \beta
\end{equation}
where
\begin{equation}
\beta = \langle D_s/D_{ds} \rangle
\end{equation}
with $D_d$, $D_{ds}$ and $D_s$ being the angular diameter distances
between observer and lens, lens and source and observer and source
respectively.

The CDFS galaxy sample selected here has a distribution of redshifts,
and hence for a given lens redshift, the value of
$\Sigma_{\mathrm{crit}}$ is a weighted average over the distribution.
At faint magnitudes the actual distribution is unknown, but we can
illustrate the effect of the differences between the redshift
estimators discussed above by calculating $\Sigma_{\mathrm{crit}}$ for
each of the estimated distributions.  We choose as test cases three
lens redshifts ($z=0.3$, 0.5 and 0.7) and use the redshift
distributions obtained from: (i) MEV estimates using a non-Bayesian
approach; (ii) MEV with Bayesian prior; (iii) full posterior
distributions; (iv) the prior distribution alone (based only on the
magnitude limits with no colour information).

\begin{table}
\begin{center}$17<R<25.5$\end{center}
\small{
\begin{tabular}{|c|c|c|c|c|}
\hline
\textbf{$z$} & \textbf{MEV} & \textbf{MEV} & \textbf{Full} &
\textbf{Prior} \\
 & \textbf{non-Bayesian} & \textbf{Bayesian} &
\textbf{posterior} & \textbf{alone} \\
\hline
0.3 & 9.96 & 11.87 & 10.78 & 9.56 \\ 
0.5 & 4.02 & 5.04 & 4.96 & 4.67 \\
0.7 & 2.01 & 2.48 & 2.71 & 3.05 \\
\hline
\end{tabular}
}
\begin{center}$17<R<23$\end{center}
\small{
\begin{tabular}{|c|c|c|c|c|}
\hline
\textbf{$z$} & \textbf{MEV} & \textbf{MEV} & \textbf{Full} &
\textbf{Prior} \\
 & \textbf{non-Bayesian} & \textbf{Bayesian} &
\textbf{posterior} & \textbf{alone} \\
\hline
0.3 & 7.83 & 7.51 & 7.79 & 7.49 \\ 
0.5 & 3.39 & 3.30 & 3.60 & 4.21 \\
0.7 & 5.03 & 4.57 & 4.11 & 3.74 \\
\hline
\end{tabular}
}
\caption{Critical surface mass densities in two magnitude ranges for
  varying lens redshifts and distribution estimates, $10^8 M_\odot
  h \mathrm{kpc}^{-2}$}
\label{critdensities}
\end{table}

For the CDFS sample with $17<R<25.5$ we find values of critical
density shown in Table~\ref{critdensities}.  At low lens redshifts the
differences between the densities are not dramatic (as is expected as
$\beta$ changes most near the lens where the estimates all largely
agree), but at higher lens redshifts the differences can be larger,
most notably between the non-Bayesian estimates and the three other
estimates at a lens redshift of $0.7$, where the non-Bayesian estimate
differs from the others by between $20\%$ and $50\%$. In addition,
although the Bayesian MEV estimates look generally reasonable, the
fact that a large number of objects that other estimates place in
front of the lens are classified instead as background objects could
lead to these ``misclassified'' objects diluting the shear signal and
causing the lens mass to be underestimated, despite getting an
apparently reasonable estimate of $\Sigma_\mathrm{crit}$ in the above
calculation (such effects are investigated below with a Monte-Carlo
simulation).

The majority of the disagreement in the above test arises from the
$R>23$ objects, as can be seen when the critical densities for
brighter objects alone ($17<R<23$) are considered. In this case, where
the colours of the objects are the primary source of information
rather than the prior, the first three distributions all provide very
similar results, but the prior alone generally disagrees slightly. The
exception is the $z=0.7$ lens, where the background sources are all in
the tail of the redshift distribution and are close to the lens, so
that slight changes in the estimation of $n(z)$ have a dramatic effect
owing to the steep variation in $\beta$ near the lens redshift.

We can use a Monte-Carlo simulation to test the effects of poor
redshift estimation and the resulting change in mass estimates owing
to errors in individual values of $\beta$ and to foreground/background
errors. We produce a Monte Carlo simulation of COMBO-17 5-band data in
which 10,000 objects are drawn randomly from the template set and
assigned an $R$-band magnitude according to the prior
distribution. Each object has its $UBVRI$ fluxes calculated from the
template colours and the assigned $R$-band magnitude, and each of the
five fluxes is scattered assuming a Gaussian error typical for a
background-limited signal in COMBO-17.

After estimating redshifts for this simulated dataset, we have a set
of simulated objects with a `true' redshift $z_i$ for even the
faintest objects, along with an estimated photometric redshift
$z_{\mathrm{est}_i}$.  We then assign each object a convergence
\begin{equation}
\kappa_i = \frac{\Sigma}{\Sigma_\mathrm{crit}(z_i)}
\end{equation}
for a fixed surface mass density $\Sigma$, and where foreground
objects are given a convergence of zero.

 Next, we calculate an estimate of the surface mass density that would
be inferred from its convergence and estimated redshift value,
\begin{equation}
\Sigma_i = \Sigma_\mathrm{crit}(z_{\mathrm{est}_i}) \kappa_i
\end{equation}
$D_{s_{\mathrm{est}_i}} > D_d$, and with a final overall estimate of
the surface mass density $\hat{\Sigma}$ being determined as
\begin{equation}
\hat{\Sigma} = \langle \Sigma_i w_i \rangle / \langle w_i \rangle .
\end{equation}
The weighting $w_i$ is a foreground/background parameter and is 1 in
the case of MEV estimates where $D_s>D_d$ and 0 elsewhere, and in the
case of a posterior distribution or a prior is given by
\begin{equation}
w_i = \int_{z_d}^{z_\mathrm{max}} p(z) dz .
\end{equation}

Resulting mass estimates in each case are given in
Table~\ref{sim_lens} for lenses at $z=0.3$, 0.5, 0.7 and 0.9. The
prior alone performs much better in this case as the `true' redshift
distribution is drawn directly from it. The unusually good result for
the non-Bayesian MEV estimates at $z=0.5$ may be due to various
effects of misestimations cancelling out by chance. The dramatically
poor result for the $z=0.9$ lens for these estimates suggest that this
choice would be a poor one in general. Even the Bayesian MEV estimates
show significant misestimations of the surface mass density and can't
therefore be recommended as an approach unless the problematic fainter
objects are excluded (at $R<23$ the Bayesian MEV estimated densities
are all within $2\%$ of the true value except for a lens redshift of
$0.9$, whereas the non-Bayesian MEV estimated densities deviate by
more than $2\%$ for all the lens redshifts).

\begin{table}
\small{
\begin{tabular}{|c|c|c|c|c|}
\hline
\textbf{$z$} & \textbf{MEV} & \textbf{MEV} & \textbf{Full} &
\textbf{Prior} \\
 & \textbf{non-Bayesian} & \textbf{Bayesian} &
\textbf{posterior} & \textbf{alone} \\
\hline
0.3 & 1.05 & 1.02 & 1.03 & 1.01 \\
0.5 & 1.01 & 1.12 & 1.01 & 0.99 \\
0.7 & 1.17 & 1.22 & 1.03 & 0.99 \\
0.9 & 1.61 & 1.23 & 1.01 & 1.04 \\
\hline
\end{tabular}
}
\caption{Surface mass densities (ratio to true value) for four
  distribution estimates and four lens redshifts from simulated data,
  $17<R<25.5$.}
\label{sim_lens}
\end{table}

When the reconstructed mass densities are dramatically incorrect they
overestimate. This is understandable if the numbers at the highest
redshift are underestimated, as is seen in the MEV estimations at low
signal-to-noise in Fig.~\ref{distributions}. Galaxies in this region
have their redshifts underestimated, and as a result the calculated
lensing mass must be higher to produce the same lensing effect at a
lower distance from the lens.

For the faintest galaxies in a real survey however, the shape
measurements will also be noisy, and will be downweighted, which will
generally reduce the effect of photometric redshift errors. This
reduction in the likely errors compared to the Monte Carlo simulations
given here will not be of as much benefit for those lensing systems at
high redshift, since the background sources in this case will all tend
to be fainter and have largely comparable weightings.

We return briefly to the issue raised in Section~\ref{priorerrors},
where the effect of noise on the measurements of the GLFs was
considered. Using the same priors as used in that section ($\alpha$
varied by $\pm 1 \sigma$ and $\phi^\star(z=1.1)$ for red galaxies
varied by $\pm 1 \sigma$) we recalculate the surface mass densities of
Table~\ref{sim_lens}. The results are given in
Table~\ref{sim_lens_altered}. The differences are still small (a few
tenths of a percent difference in surface mass densities between the
various priors) and whilst the error on the prior may be larger and
the inaccuracy in the prior may be large enough to have appreciable
effects, they are still likely to be much smaller than those effects
that arise from a non-Bayesian or MEV estimation approach.

\begin{table}
\small{
\begin{tabular}{|c|c|c|c|c|c|}
\hline
\textbf{$z$} & \textbf{Unaltered} & $\alpha+1\sigma$ & $\alpha-1\sigma$ &
$\phi^\star+1\sigma$ & $\phi^\star-1\sigma$ \\
\hline
0.3 & 1.028 & 1.029 & 1.028 & 1.029 & 1.029 \\
0.5 & 1.014 & 1.015 & 1.014 & 1.015 & 1.014 \\
0.7 & 1.026 & 1.030 & 1.028 & 1.031 & 1.027 \\
0.9 & 1.008 & 1.009 & 1.011 & 1.011 & 1.009 \\
\hline
\end{tabular}
}
\caption{Surface mass densities (ratio to true value) for full posterior
  distribution estimates and four lens redshifts from simulated data,
  $17<R<25.5$, using priors calculated from altered GLF parameters.}
\label{sim_lens_altered}
\end{table}

We also return to the issue of biases in redshift estimates and
issues of higher order moments of the posterior probability
distribution. Section~\ref{usingfullprior} gave estimates of the bias
of photometric redshifts, expressed as $\langle 1+z_{\mathrm{photo}}
\rangle = b(1+z_{\mathrm{spec}})$. Fitting to the VVDS sample at
$0.1<z<1.0$ gave $b=0.978$, which would lead to estimated redshifts
generally falling slightly below the true redshift, and therefore lead
to a slight overestimation in lensing masses. In the case of a single
source at $z=1$, for lens redshifts of 0.3, 0.5 and 0.7 the
misestimations are $1.2\%$, $2.8\%$ and $6.7\%$ respectively, much
lower than those arising from choosing an alternative estimation
technique (although this bias was measured only for the restricted
range of objects described earlier, and will be more of a problem at
$z>1$).

Higher order moments of the posterior probability are also a concern,
since the critical densities we estimate are based on $\beta = \langle
D_s/D_{ds} \rangle$, and $\langle \beta \rangle$ is not proportional
to $\langle z \rangle$.

However, since $\beta$ approaches a constant as $D_s$ increases,
variations in the distribution at high redshift do not impact the
estimated lensing mass as much as estimates at low redshift. As a
result, for low lens redshifts the shape and width of the
distributions are not a major concern, since close to the lens where
$\beta$ is varying most rapidly the sources are also at low redshift and,
being generally brighter, have tightly constrained distributions. At
high lens redshifts the distribution widths and shapes are much more
of a concern as they will be broader and cover regions of significant
variation in $\beta$. In these cases estimates of the bias alone may
prove insufficient and the best approach to estimating resulting
errors would likely be further simulations.

The conclusion from these tests is that the full posterior approach is
capable of providing lensing estimates that agree with previously used
photometric techniques in the $R<23$ regime, and yet also agrees for
fainter samples with the modelled redshift distributions, providing
consistent estimated lens masses over a broad range of lens redshifts.
In these tests Bayesian MEV estimates perform better than pure
likelihood estimates, but are likely to be less good in 3-D lensing
analyses.  In such analyses, the distribution of mass is estimated as
a function of redshift using the photometric redshifts of the observed
galaxies.  Hence we expect the estimated mass distribution will be
even more sensitive to any systematic poor estimation of redshifts
than the tests that we have described above.

\section{Conclusions}
Extraction of the full information content of future generations of
weak lensing surveys will require extensive use of photometric
redshifts, applied to both 2D and 3D analyses.  Weak lensing signals
are concentrated towards the magnitude limits of surveys, and yet it
is here that photometric measurement errors make photometric redshift
estimation the most unreliable.  These errors lead not only to a broad
distribution of redshift uncertainty, but also to increasing severity
of the degeneracy between galaxy type and redshift that frequently
leads to multiple solutions for an estimated redshift.  Furthermore,
the effect of measurement errors is to make the sample distribution of
estimated redshifts different from the true distribution of redshifts,
which inevitably leads to bias in the values of cosmological
parameters estimated from the overall sample.

In this paper we have considered the Bayesian solution to this
problem, and have shown that using the full Bayesian posterior
probability distribution for estimated redshifts allows such biases to
be eliminated, provided a prior distribution, here based on knowledge
of the galaxy luminosity function and its evolution, is known. Using
photometric and redshift data from the COMBO-17 sample, we have
compared the full Bayesian approach with pure likelihood estimation
and minimum error variance (MEV) methods and show that, although there
is little to choose between methods at bright magnitudes, the full
Bayesian method is significantly less biased at faint magnitudes.
Furthermore, the Bayesian method allows the sample distribution of
estimated redshifts to tend smoothly towards the prior distribution at
the faintest limits of a survey.

Applying the method to simulated weak lensing signals, we show that
reconstruction of lensing mass density in the presence of photometric
redshift errors can lead to biases in excess of 20 percent in 2D
lensing analyses unless the Bayesian method is adopted.  We anticipate
that 3D lensing analyses will be even more susceptible to this effect,
but we postpone to a future paper a full incorporation of the Bayesian
probability approach into 3D lensing analysis.
 
\vspace*{5mm}
\noindent
{\large \bf ACKNOWLEDGEMENTS}\\ 
EME acknowledges the support of a
PPARC Graduate Studentship.  CW acknowledges the support of a PPARC
Advanced Fellowship.  We also thank David Bacon and Devinder Sivia for
many helpful comments.


\label{lastpage}

\end{document}